\documentclass[useAMS,usenatbib]{mn2e}

\usepackage{superfig,morefloats}
\usepackage{graphicx}
\usepackage{lscape}
\input{acentos.sty}

\title[Long-term optical/infrared variability in V404 Cyg]{Long-term optical/infrared variability in the quiescent X-ray transient V404 Cyg}
\author[C. Zurita, J. Casares, R.I. Hynes, T. Shahbaz, P.A. Charles and E.P. Pavlenko]{
C. Zurita$^{1}$, J. Casares$^{2}$, R. I. Hynes$^{3,4}$, T. Shahbaz$^{2}$, P. A. Charles$^{4}$ and E. P. Pavlenko$^{5}$ \\
$^{1}$Centro de Astronomia e Astrof\'{\i}sica da Universidade de Lisboa, Observat\'orio Astron\'omio de Lisboa, Tapada da Ajuda\\
1349-018 Lisboa, Portugal\thanks{E-mail:czurita@oal.ul.pt (CZ)}\\
$^{2}$Instituto de Astrof\'{\i}sica de Canarias, V\'{\i}a Lactea, E-38200 La Laguna, Tenerife, Canary Islands, Spain\\
$^{3}$Astronomy Department, University of Texas at Austin, 1 University Station C1400, Austin, Texas 78712; USA\\
$^{4}$Department of Physics and Astronomy, University of Southampton, Southampton SO17 1BJ, UK\\
$^{5}$Crimean Astrophysical Observatory, Ukraine
}

\begin{document}

\date{Accepted 1988 December 15. Received 1988 December 14; in original form 1988 October 11}

\pagerange{\pageref{firstpage}--\pageref{lastpage}} \pubyear{2002}

\maketitle

\label{firstpage}

\begin{abstract}
We  present the  results of  optical  and infrared  photometry of  the
quiescent X-ray transient V404  Cyg during the period 1992--2003.  The
ellipsoidal  modulations extracted  from the  most  complete databases
(years  1992, 1998  and  2001)  show unequal  maxima  and minima  with
relative strength  varying from  year to year  although their  peak to
peak amplitudes  remain roughly constant  at 0.24$\pm$0.01 magnitudes.
Fast   optical  variations  superimposed   on  the   secondary  star's
double-humped ellipsoidal  modulation were detected every  year with a
mean  amplitude of  $\sim$0.07\,mags.  We  have not  found significant
changes in the  activity during this decade which  indicates that this
variability is probably  not connected to the 1989  outburst.  We have
found  periodicities  in  the  1998  and  2001  data  near  the  6\,hr
quasi-periodicity  observed  in  1992,  although we  interpret  it  as
consequence of  the appearance  of a flare  event almost  every night.
Significant variability is  also present in the $I$  and near infrared
($J$  and  $K_s$)  bands   and  this  decreases  slightly  or  remains
approximately  constant  at longer  wavelengths.  A cross  correlation
analysis shows  that both  the $R$ and  $I$ emission  are simultaneous
down to 40\,s.
\end{abstract}


\section{Introduction}

Low-Mass  X-ray  Binaries (LMXBs),  are  semi-detached binary  systems
where a  normal low-mass late type  star transfers gas  onto a compact
remnant i.e. a neutron star  or black hole.  A significant fraction of
LMXBs are soft X-ray transients (SXTs) characterized by episodic X-ray
outbursts  where the  luminosity increases  dramatically.   During the
long intervals (usually many years) between outbursts, the SXTs are in
quiescence and the optical emission is dominated by the radiation from
the companion star.   This offers a superb opportunity  to analyze and
obtain dynamical information which  eventually enables us to constrain
the  nature  of  the  compact object  (see  eg.   \citealt{charles98},
\citealt{casares01}).  The majority of these systems appear to contain
black holes and  they are, in fact, the best evidence  we have for the
existence of stellar mass black holes. The most secure case is the SXT
V404  Cyg  which  has  a  mass  function  of  $6.08\pm0.06\,M_{\odot}$
(Casares, Charles  \& Naylor 1992),  twice the usually  accepted upper
limit to that of a neutron star \citep{rhoades71}.  The estimated mass
for  the  compact  object is  12$\pm$2\,M$_{\odot}$  \citep{shahbaz94}
which,  together with  GRO  1915+105 \citep{greiner01},  are the  most
massive stellar mass black holes yet measured.\\

Furthermore, V404 Cyg is considered  a peculiar SXT because it has one
the longest orbital periods  (6.5\,d; \citealt{casares92b}) and is the
most          X-ray           luminous          in          quiescence
($L_X\sim10^{33}-10^{34}$\,erg\,s$^{-1}$;  \citealt{garcia01}).  Apart
from  this, it  is largely  known  that V404  Cyg exhibits  short-term
(sub-orbital)  variations during quiescence  in the  optical continuum
\citep{wagner92,pavlenko96}         and         H$\alpha$         flux
\citep{casares92,casares93} where  the H$\alpha$ profiles  seems to be
correlated   to   the   continuum   \citep{hynes02}.    The   infrared
\citep{sanwal95},   X-rays    \citep{wagner94,   kong02}   and   radio
\citep{hjellming00}   are    also   variable.    Recent   simultaneous
observations  in  X-rays  and  optical  show  that  optical  line  and
continuum variability are well correlated with X-rays flares (Hynes et
al. in preparation).  On the other hand it seems that fast variability
during quiescence is  not a peculiarity of V404  Cyg but a fingerprint
of quiescent  SXTs, irrespective of  the nature of the  compact object
\citep[][hereafter ZCS]{zurita03,  hynes03}. Although these variations
seem to  originate in  the accretion disc  rather than in  the optical
companion or gas  stream/hot spot \citep[][ZCS]{zurita98, hynes02} the
underlaying physical  mechanism is not yet  understood.  The dynamical
(Keplerian) timescales associated with these fluctuations suggest that
the  flares are  produced in  the outer  disc (ZCS).   Alternatively ,
based  on the  similarities of  the  power density  spectrum (PDS)  of
A0620-00  to  those of  low/hard  state  sources \citet{hynes03}  have
pointed out that the optical variability is at least partly related to
the inner accretion flow even  in quiescence. This suggestion was also
supported by the discovery of  the correlated optical and X-ray flares
(Hynes et al. in preparation).\\

Since V404  Cyg is  the brightest SXT,  and hence accessible  with 1-m
type telescopes, we  embarked on a study of  the fast variability over
the period 1992--2003. This  paper comprises of a simultaneous optical
and infrared photometric study, with the aim to determining the colour
of the variability and its possible origin.

\section{Observations and Data Reduction}

We  obtained  CCD  photometry  of  V404 Cyg  with  various  1\,m-class
telescopes during the years 1992, 1998, 1999, 2001, 2002 and 2003. The
1992  data were  obtained in  the Johnson  $R$ band  with the  GEC CCD
camera  on the 1\,m  JKT telescope  at Observatorio  del Roque  de los
Muchachos.  V404  Cyg was observed  in 1998 simultaneously in  the $R$
and $I$  band with  identical Thomson CCD  cameras at both  the 0.8\,m
IAC80  and 1\,m  OGS telescopes  at Observatorio  del Teide.   We also
obtained $R$ band photometry in 1999  with the SITe2 CCD camera on the
JKT. Further details on the 1992 and 1999 observations can be found in
\citet{pavlenko96} and  \citet{hynes02} respectively.  During Jul--Sep
2001  $R$-band  and  $J$  and  $K_s$  simultaneous  observations  were
performed using  the Thomson  CCD camera at  the 0.8\,m IAC80  and the
infrared camera at the  1.5\,m {\em Telescopio Carlos S\'anchez} (TCS)
also at Observatorio  del Teide.  Finally we observed  V404 Cyg during
May--Jun 2002  and Jul  2003 also in  the $R$-band and  with identical
Thomson CCD cameras  at both the 0.8\,m IAC80  and 1\,m OGS telescopes
at  Observatorio del  Teide. Full  details of  these  observations are
given in Table~\ref{log}.\\

All images  were corrected for  bias and flat-fielded in  the standard
way  using {\sc  iraf}.  Seeing  conditions  were not  good enough  to
cleanly  separate  the  contribution  of  the target  and  its  nearby
(1.4\,arcsec)  line-of-sight  star  \citep{udalski91}  so  we  applied
straightforward  aperture   photometry  using  a   large  aperture  of
3.5\,arcsec  which  adds  the  flux  from both  stars  and  which  was
subsequently corrected  (see next section).   Several comparison stars
within  the field  of view  were checked  for variability  during each
night and during the entire data set.

\begin{table*}
 \centering
 \begin{minipage}{140mm}
  \caption{Log of observations}
\label{log}
  \begin{tabular}{@{}lccc@{}}
  \hline
{\em Night} & {\em Band} & {\em Resolution~(s)} & {\em Monitoring time~(hr)}\\
 \hline
\multicolumn{4}{l}{\bf 1992}\\
\multicolumn{4}{l}{1\,m JKT}\\
{Jun 27} & {$R$}  & {259} & {1.45}\\
{Jun 28} & {$R$}  & {259} & {5.44}\\
{Jun 29} & {$R$}  & {259} & {2.51}\\
{Jun 30} & {$R$}  & {259} & {7.11}\\
{Jul 1} & {$R$}  & {259} & {7.07}\\
{Jul 2} & {$R$}  & {259} & {5.33}\\
{Jul 3} & {$R$}  & {259} & {6.18}\\
{Jul 4} & {$R$}  & {259} & {5.15}\\
{Jul 5} & {$R$}  & {259} & {4.11}\\
{Jul 6} & {$R$}  & {259} & {6.16}\\
{Jul 7} & {$R$}  & {259} & {6.37}\\
{Jul 8} & {$R$}  & {259} & {6.10}\\
{Jul 9} & {$R$}  & {259} & {7.24}\\
{Jul 10} & {$R$}  & {259} & {5.67}\\
{Jul 11} & {$R$}  & {259} & {6.68}\\
{Jul 12} & {$R$}  & {259} & {6.29}\\
\noalign{\smallskip}
\multicolumn{4}{l}{\bf 1998}\\
\multicolumn{4}{l}{0.8\,m IAC80\,/\,1\,m OGS}\\
{17 Aug} & {$R$}  & {94} & {3.98}\\
{18 Aug} & {$R$}  & {88} & {6.09}\\
{19 Aug} & {$R$}  & {79} & {6.32}\\
{21 Aug} & {$R$\,/\,$I$}  & {129\,/\,99} & {6.52\,/\,7.38}\\
{22 Aug} & {$R$\,/\,$I$}  & {146\,/\,99} & {6.21\,/\,5.88}\\
{23 Aug} & {$R$\,/\,$I$}  & {136\,/\,129} & {6.77\,/\,1.08}\\
{24 Aug} & {$R$\,/\,$I$}  & {136\,/\,129} & {7.03\,/\,7.28}\\
{25 Aug} & {$R$\,/\,$I$}  & {138\,/\,129} & {6.74\,/\,7.21}\\
{26 Aug} & {$R$}  & {136} & {4.54}\\
{27 Aug} & {$R$}  & {136} & {7.19}\\
\noalign{\smallskip}
\multicolumn{4}{l}{\bf 1999}\\
\multicolumn{4}{l}{1\,m JKT}\\
{Jul 6} & {$R$}  & {80} & {5.89}\\
{Jul 7} & {$R$}  & {80} & {6.05}\\
\noalign{\smallskip}
\multicolumn{4}{l}{\bf 2001}\\
\multicolumn{4}{l}{0.8\,m IAC80\,/\,1\,m TCS}\\
{31 Jul 01} & {$R$}  & {314}  & {3.80}\\
{1 Aug} & {$R$}  & {314}  & {4.12}\\
{2 Aug} & {$R$}  & {314}  & {5.22}\\
{3 Aug} & {$R$\,/\,$J$\,/\,$K_s$}  & {314\,/\,1062\,/\,1100} & {4.98\,/\,4.80\,/\,2.56}\\
{4 Aug} & {$R$}  & {314} & {4.17}\\
{6 Aug} & {$R$\,/\,$J$\,/\,$K_s$}  & {314\,/\,1054\,/\,1100} & {3.88\,/\,1.80\,/\,3.72}\\
{7 Aug} & {$R$\,/\,$J$\,/\,$K_s$}  & {314\,/\,1030\,/\,1040} & {3.37\,/\,3.05\,/\,2.73}\\
{8 Aug} & {$R$\,/\,$J$\,/\,$K_s$}  & {314\,/\,1030\,/\,1050} & {4.04\,/\,2.53\,/\,3.83}\\
{10 Aug1} & {$R$\,/\,$J$\,/\,$K_s$}  & {314\,/\,1630\,/\,1050} & {4.34\,/\,4.44\,/\,3.53}\\
{11 Aug1} & {$R$}  & {197} & {3.87}\\
{28 Aug1} & {$R$\,/\,$J$\,/\,$K_s$}  & {314\,/\,1100\,/\,1050} & {5.19\,/\,5.32\,/\,5.05}\\
{30 Aug1} & {$R$\,/\,$J$\,/\,$K_s$}  & {314\,/\,1150\,/\,1050} & {5.94\,/\,5.36\,/\,5.04}\\
{31 Aug1} & {$R$}  & {314} & {6.54}\\
{6 Sep} & {$R$}  & {314} & {3.84}\\
{8 Sep} & {$R$}  & {314} & {4.78}\\
\hline
\end{tabular}
\end{minipage}
\end{table*}

\begin{table*}
\addtocounter{table}{-1}
 \centering
 \begin{minipage}{140mm}
  \caption{Log of observations (continuation)}
\label{log}
  \begin{tabular}{@{}lccc@{}}
  \hline
{\em Night} & {\em Band} & {\em Resolution~(s)} & {\em Monitoring time~(hr)}\\
 \hline
\multicolumn{4}{l}{\bf 2002:}\\
\multicolumn{4}{l}{1\,m OGS}\\
{27 May 02} & {$R$}  & {315}  & {4.02}\\
{28 May 02} & {$R$}  & {315}  & {3.90}\\
{02 Jun 02} & {$R$}  & {315}  & {4.27}\\
{03 Jun 02} & {$R$}  & {315}  & {4.42}\\
\multicolumn{4}{l}{0.8\,m IAC80}\\
{01 Jun 02} & {$R$}  & {315}  & {3.37}\\
{04 jun 02} & {$R$}  & {315}  & {2.05}\\
\noalign{\smallskip}
\multicolumn{4}{l}{\bf 2003:}\\
\multicolumn{4}{l}{1\,m OGS}\\
{17 Jul 03} & {$R$}  & {252}  & {1.20}\\
{18 Jul 03} & {$R$}  & {252}  & {8.20}\\
{19 Jul 03} & {$R$}  & {252}  & {4.33}\\
{20 Jul 03} & {$R$}  & {252}  & {8.26}\\
\multicolumn{4}{l}{0.8\,m IAC80}\\
{28 Jul 03} & {$R$}  & {352}  & {8.72}\\
\hline
\end{tabular}
\end{minipage}
\end{table*}

\section{The $R$ band light curves}
\label{curvas_v4R}

\subsection{The ellipsoidal modulation}

The true V404 Cyg $R$-band lightcurves were obtained after subtracting
the contribution  of the {\it contaminating}  companion star, assuming
that it had $R$=17.52$\pm$0.01  as given in \citet{casares93}. We then
folded  our  data using  the  ephemeris  of \citet{casares94}.  These
lightcurves show the classical secondary's ellipsoidal modulation (see
\citealt{shahbaz94}) although with superposed variability.\\

\begin{figure}
\includegraphics[width=8cm]{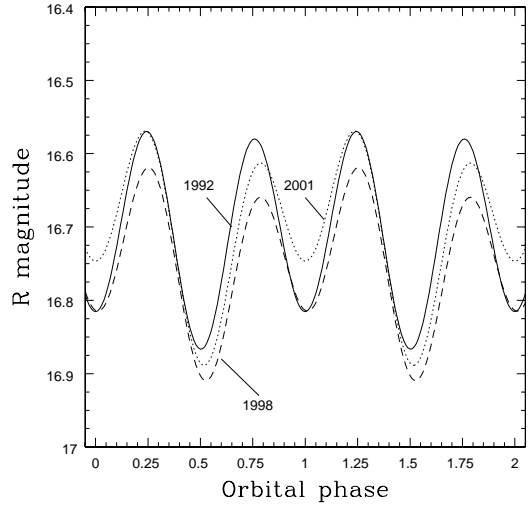}
\caption{Ellipsoidal  model fits to  the lower  envelopes of  the 1992
(continuum line), 1998 (dashed  line) and 2001 (pointed line) $R$-band
lightcurves.}
\label{elipmodel}
\end{figure}

The ellipsoidal  modulation is due  to the differing aspects  that the
tidally distorted secondary star  presents to the observer through the
orbital cycle.  An ideal pure ellipsoidal lightcurve is hence a double
humped  modulation with two  equal maxima  and two  somewhat different
minima (the minimum  at phase 0.5 is deeper  because gravity darkening
is   more   pronounced  in   the   direction   of   the  $L_1$   point
\citep{avni75}. This modulation was first simulated employing two sine
waves  at the  orbital frequency  and  its first  harmonic, where  the
fundamental allows the minima to  vary.  Relative phases were fixed to
produce  equal maxima.   However, it  has been  largely  reported that
lightcurves of  quiescent transients  often exhibit unequal  maxima as
well   (eg.     A0620-00:   \citealt{haswell96,gelino01}-   J1118+480:
\citealt{zurita02}- J2123-058:  \citealt{shahbaz03}).  The distortions
from the  theoretical case could be explained  assuming a non-constant
contribution  of  the accretion  disc,  X-ray  heating,  spots on  the
surface  of  the  secondary  star  or light  from  the  stream  impact
point. In any  case, we also account for  this possibility by allowing
the  sine phases  to vary  independently.   We then  fitted these  two
models to  the mean light curve  and also to the  lower envelope.  The
lower  envelope was constructed  as has  been previously  explained in
\cite{pavlenko96} and  ZCS, that is,  for each night we  have selected
all brightness estimates which differ from the minimum by no more than
twice the accuracy  of the observations.  The selected  data were then
averaged   in  time   and  phase   intervals.   We   found   the  best
representation (the minimum $\chi^2$)  fitting the lower envelope with
the  second model.   These fits  for the  years with  a  most complete
database (1992,  1998 and  2001) are shown  in Figure~\ref{elipmodel}.
Both minimum and maximum magnitudes  varies. The maximum at phase 0.25
is higher  than at  phase 0.75 and  also the  minimun at phase  0.5 is
deeper than at  phase 0. However their relative  strength changes from
year to  year.  Apart  from this,  the peak to  peak amplitude  of the
first harmonic remains roughly constant at 0.24$\pm$0.01 magnitudes, a
key fact  in the  modeling of  V404 Cyg and  the determination  of its
component  masses.   We note  that  the  1998  lightcurve obtained  by
\cite{pavlenko01} in  the same band exhibits equal  maxima, however we
cannot conclude whether  this behaviour reflects a real  change in the
system, since both  lightcurves are not simultaneous, or  is due to an
incomplete sampling.

\subsection{The superimposed fast variability}

\begin{figure}
\includegraphics[width=8cm]{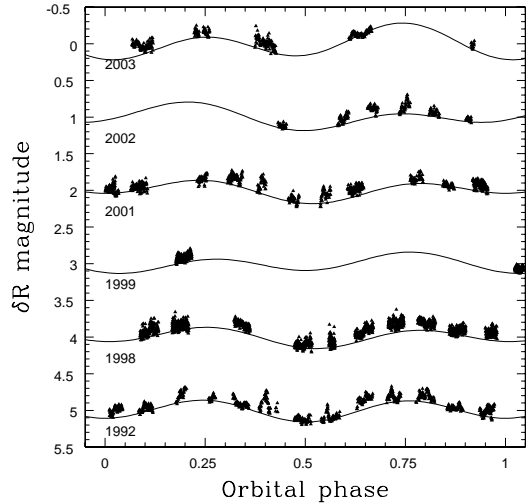}
\caption{The $R$  band ellipsoidal lightcurves  of V404 Cyg  for 1992,
1998, 1999, 2001, 2002 and 2003 and the lower envelopes constructed as
explained  in section  \ref{curvas_v4R}.  Different magnitude  offsets
were aplied for clarity.}
\label{elip+var}
\end{figure}

We assumed  that the fast variability superimposed  on the ellipsoidal
curve is not  related to the ellipsoidal modulation  so the former was
isolated by  subtracting the lower  envelope variable-phase sinusoidal
fit  (as has  been  explained  in the  previous  subsection) from  the
original data  resulting in  a light curve  essentially free  from the
ellipsoidal modulation.   In the case of the  1999 lightcurve, because
only  two nights  of observations  were  available, we  fixed all  the
parameters of the  double sine wave obtained from  the 1998 fit except
the phase of the first sine.  In 2003 only five nights of observations
are available so we fixed the mean magnitude in the variable-phase fit
at the  value we  obtained with the  simplest constant  relative phase
fit.   Furthermore, in  order to  compare the  results  from different
years, we  have resampled  the lightcurves to  have the  same temporal
resolution  (320\,s).   The lightcurves  and  the corresponding  lower
envelope fits are shown in Figure~\ref{elip+var} and the result of the
subtraction is represented in Figure~\ref{var}.\\

Examining  these  lightcurves we  find  that  the observed  variations
resemble  the lightcurves of  stellar flares,  i.e.  rapid  and strong
increases  in  brightness.   We  consider  a {\it  flare}  to  be  any
variation which differs from the mean level by more than 3$\sigma$. We
found  in  many nights  large  single  events  of $\sim$2\,h  or  more
duration   appearing  together   with  faster   and   small  amplitude
variations.   Other nights,  however, only  the faster  variations are
present.  The shape  of the  large flares  seem to  be  symmetric with
similar rise and decay times.  In order to characterize the flares and
quantify the flare activity we calculate the mean, $\bar{z}$, standard
deviation corrected for the instrumental error, $\sigma_{z}$ (i.e.  we
subtract the  instrumental error in  quadrature ) and  {\it equivalent
duration}  ($\Pi_{z}$)  of the  relative  intensity lightcurves,  i.e.
relative to the steady non-variable level (see ZCS).

Here {\it z} is the {\it relative intensity}, defined as: 
\begin{equation}
z= (I-I_{\rm q})/I_{\rm q}
\end{equation}
where $I$ and $I_{\rm q}$ are the observed intensities of the flaring and not flaring state respectively.

The equivalent duration ($\Pi$) is defined as:

\begin{equation}
\label{eq_duration}
\Pi  = \int{z\,d\,t} \hspace{3mm}
\end{equation}

\noindent 

The total energy  of the flare  ($E$)  is obtained by multiplying  the
equivalent duration by the  system's luminosity.  i.e.  $E=4\,\pi\,d^2
F_q\,\Pi$\,erg  (where $d$ is the distance  to the system and $F_q$ is
the  dereddened  observed  quiescent  flux).  The time-averaged  flare
luminosity $<L_{R}>$ is  defined  as the sum   of the energies of  the
individual flares   divided by  the  total monitoring   time  ($M_{\rm
t}$). The system   properties  and the  resulting parameters  of   the
variability are       summarized       in   Tables~\ref{propiedades_v4}
and~\ref{variabilidad_v4} respectively.\\

\begin{table*}
 \centering
\begin{minipage}{140mm}
  \caption{V404 Cyg properties.}
  \label{propiedades_v4}
  \begin{tabular}{@{}ll@{}}
  \hline
{\em Distance (kpc)} &  3.5$\pm$0.2\\
{\em $A_V$} &  4.0\\
{\em Spectral Type} &  K0{\sc iv}\\
{\em inclination} &  56$^{\circ}\pm$2\\
{\em P$_{orb}$ (days)} & 6.64714\\
{\em Dereddened magnitudes} & $R$=13.5, $I$=13.81, $J$=12.55, $K$=12.05 \\
\hline
\end{tabular}

\medskip
Parameters extracted from \citet{wagner92};
\citet{shahbaz94}; \citet{pavlenko96}; \citet{martin99} 
and \citet{casares93}.
\end{minipage}
\end{table*}

\begin{table*}
 \centering
\begin{minipage}{140mm}
  \caption{V404 Cyg flares properties for the different years.}
  \label{variabilidad_v4}
  \begin{tabular}{@{}lccccc@{}}
\hline
{\em  Year}  &  {\em  $<\Pi^{\dagger}>$} &  {\em  Monitoring}  &  {\em
$<L_{R}>$} &  {\em $\bar{z_f}$} & {\em $\sigma_z$}  \\ 
             & 	                                & {\em  time (hr)}    &  {\em $\times$10$^{32}$\,erg\,s$^{-1}$} &  & \\
  \hline
1992 & 0.055 & 88.9 & 10.7$\pm$0.4 & 0.057 & 0.040\\
1998 & 0.084 & 61.4 & 15.8$\pm$0.5 & 0.084 & 0.034\\
1999 & 0.066 & 11.9 & 12.5$\pm$0.4 & 0.067 & 0.025\\
2001 & 0.065 & 68.1 & 12.7$\pm$0.4 & 0.061 & 0.042\\
2002 & 0.071 & 22.0 & 15.1$\pm$0.5 & 0.064 & 0.038\\
2003 & 0.073 & 30.7 & 13.8$\pm$0.5 & 0.067 & 0.041\\
\hline
\end{tabular}

\medskip
($\dagger$) Time averaged equivalent duration of the individual
flares divided by the total monitoring time.
\end{minipage}
\end{table*}

We  note  that  the  flare  mean level  and  the  time-averaged  flare
luminosity are  somewhat sensitive to  the ellipsoidal model  which is
subtracted.  It may  be that the model is  not an ideal representation
of the non-variable  level due to our incomplete  sampling and effects
other than  the flares which  distort the ellipsoidal  modulation.  In
contrast,  the standard  deviation ($\sigma_z$)  is  not significantly
affected by the  model assumed.  We hence consider  that $\sigma_z$ is
the best  flare level  activity indicator.  We  do not find  any clear
trend towards increasing or  diminishing flare activity with time (see
Table~\ref{variabilidad_v4}  and  Fig.~\ref{sample}).   Moreover,  the
long-term photometric study of V404 Cyg during quiescence performed by
\citet{pavlenko01} also shows the fast variability persisting over the
decade 1992-2000  with variable amplitude.  In particular  in 2000 the
mean amplitude was 0.06.  This is consistent with our data as is clear
in Fig.~\ref{sample}.   We estimate the  flare activity to  be roughly
constant    with     $\bar{z}\sim$0.065,    $\sigma_z\sim$0.036    and
$<L_{R}>\sim$13$\times$10$^{32}$\,erg\,s$^{-1}$.   Assuming  that  the
X-ray              (0.3--7\,kev)             luminosity             is
$<L_{X}>\sim$5$\times$10$^{33}$\,erg\,s$^{-1}$   (from   the   Chandra
spectrum;  \citealt{kong02}) we  estimated the  ratio  between optical
flare and X-rays luminosities: $<L_{opt}/L_{X}>\sim$0.3

\section{Searching for Periodicities}
\label{periodos_v4}
As the short  term variations are non-sinusoidal we  have attempted to
analyze and separate the periodicities  present in the data by fitting
a  Fourier series  to the  lightcurves  over a  range of  frequencies,
$\nu$ \citep{martinez94}. A plot of  the reduced $\chi^2$ of each fit
versus $\nu$  should have  a minimum at  the fundamental  frequency of
oscillation in the data.\\

We applied this method after  resampling the data from different years
to  the same  time  resolution, and  after  detrending the  long--term
variations  by subtracting  the  lower envelope  as  explained in  the
previous section.   We rejected those  nights having less  than 3\,hrs
coverage so  as to  avoid peculiar offsets  due to poor  sampling. The
$\chi^2$ spectrum  is computed in  the range 3 to  10 cycles\,d$^{-1}$
with a  resolution of  5$\times$10$^{-2}$ cycles\,d$^{-1}$ and  with 3
terms  in the  Fourier series.   They are  shown  in Fig.~\ref{period}
where  we  have  also  marked  the  68  percent  confidence  limit  at
$\chi^2_{min}+1$ \citep[e.g.,][]{lampton76}\\

With  the 1992  data, we  reproduce the  results already  published in
\citet{pavlenko96}.   There exist  a  peak in  the  vicinity of  6\,hr
(0$^d$.25) which is split into two (at 0$^d$.260 and 0$^d$.246) with a
central peak  coinciding with  the median value  of these  periods (at
0$^d$.253).  The 1\,day aliases  are also visible. A detailed analysis
of these data  can be found in \citet{pavlenko96}.   The 1998 spectrum
shows two  peaks at 0$^d$.247  and 0$^d$.233 exceeding the  68 percent
confidence interval and also their 1\,day aliases.  The power spectrum
of the  2001 data is very noisy  and we found two  possible periods at
0$^d$.241 and 0$^d$.235 (also with  their aliases) and two other peaks
at 0$^d$.261 and 0$^d$.250 all of them with the same significance.  We
note that these  data were also analysed in  ZCS.  Here a Lomb-Scargle
periodogram was performed yielding significant peaks only above 1 hour
but  with a  complicated alias  pattern  which makes  it difficult  to
distinguish   any   periodicity.    Finally,   we   also   note   that
\citet{hynes02} analysis  of the 1999 lightcuves obtained  at CrAO and
SAI \footnote{with  the 0.38-m telescope of  the Crimean Astrophysical
Observatory (CrAO) and the 1.25- and 0.6-m telescopes of the Sternberg
Astronomical  Institute (SAI) in  1999.}  together  with our  1999 JKT
data could not find any clear periodicity.  Here the peak with highest
significance has a period of 0$^d$.254 although other periods are also
possible.  Clearly  there is  no coherent period  in our  database but
variability on a characteristic timescale of about 6 hours is detected.\\

\section{The colour of the flares}

\subsection{The 1998 simultaneous $R$ and $I$ photometry}

V404 Cyg  was also  observed simultaneously in  the $R$ and  $I$ bands
during  1998.  To  extract the  true $I$  magnitudes of  V404  Cyg, we
previously  subtracted  the contribution  of  the {\it  contaminating}
star, $I_{cs}$.   We calibrated the  unresolved pair of V404  Cyg plus
contaminating   star   using   the   $I$   magnitudes   tabulated   in
\citet{martin99} and then subtracted the real median magnitude of V404
to  find $I_{cs}=17.0\pm$0.1.   The resulting  lightcurve is  shown in
figure~\ref{elipI}, where fast  variability is clearly superimposed on
the  ellipsoidal  modulation.   We  detrended the  long--term  ellipsoidal
modulation  by   subtracting  the  lower  envelope   as  explained  in
section~\ref{curvas_v4R}. We also binned the lightcurves to 300\,s.\\

Variability is  clearly seen  in the $I$  band as well,  following the
same   pattern   as   the   variability   in   the   $R$   band   (see
Fig.~\ref{curvaRI}). We calculate the  mean, standard deviation of the
lightcurves and the equivalent duration  of the flares relative to the
steady non-variable level in both the $R$ and $I$ bands.  We note that
although the time  resolution is the same, the  parameters for the $R$
band change  from those reported  in section~\ref{parametros_v4} since
now  we   are  only  including   in  our  analysis  the   nights  with
silmultaneous    observations.     The    results   are    given    in
table~\ref{flares_RI}.\\

\begin{table*}
 \centering
\begin{minipage}{100mm}
  \caption{V404 Cyg flares properties for the $R$ and $I$ bands in 1998.}
  \label{flares_RI}
  \begin{tabular}{@{}lccccc@{}}
\hline
{\em  Band}  &  {\em  $<\Pi^{\dagger}>$} &  {\em  Monitoring}  &  {\em
$<L>$} &  {\em $\bar{z_f}$} & {\em $\sigma_z$}  \\ & & {\em  time (hr)} &
{\em $\times$10$^{32}$\,erg\,s$^{-1}$} & & \\
  \hline
R & 0.097 & 28.8 & 18.2$\pm$0.6 & 0.095 & 0.030\\
I & 0.038 & 28.8 &  3.8$\pm$0.1 & 0.039 & 0.019\\
\hline
\end{tabular}

\medskip
($\dagger$) Time averaged equivalent duration of the individual
flares divided by the total monitoring time.
\end{minipage}
\end{table*}

The flare  activity in  the $R$ band  is larger  than in the  $I$ band
although we note that the  relative contribution of the secondary star
and  the  accretion disc  to  the total  light  is  different at  both
wavelengths. In the  case of V404 Cyg the  distribution of the veiling
rises   toward   the   blue   following   a   power   law   of   index
$\alpha=-3.4\pm0.9$   which  basically  reflects   the  drop   in  the
photospheric flux  of the last-type companion short  ward of 5300\,\AA
\citep{casares93}. We also calculated the fluxes of the flares in both
bands. To do  it we first de-redden the  observed magnitudes using for
the reddening a value of $A_V$=2.8 (see \citealt{shahbaz03b}) and then
subtracted the ellipsoidal modulation  constructed as explained in the
previous  sections  from   the  de-reddened  lightcurve  yielding  the
monochromatic  flux of  the  flares. We  obtained $F_R=1.25\,mJy$  and
$F_I=0.41\,mJy$,  so the  mean flux  density  ratio of  the flares  is
$F_R/F_I\sim3$. Nevertheless there exists  several sources of error in
the calculation of  the flare fluxes. First, there  is the photometric
errors as  well as  the correction  of the flux  from the  nearby star
which  contaminates  V404  Cyg  (see section  \ref{curvas_v4R}).   The
photometric  accuracy  is  $\sim$1\,percent  in  both  bands  but  the
magnitude  of the  contaminating star  is known  with a  $\sim$1\% and
$\sim$10\% accuracy in $R$ and  $I$ respectively although, in fact, we
must consider  these values as  lower limits since they  were obtained
from  the literature and  there is  difference between  effective band
passes on  different telescopes. But the  main source of  error is the
determination of the reddening.  If  we use the extreme values for the
reddening A$_V$=4.0 and A$_V$=2.2 \citep{casares93,shahbaz94} we found
an  80\%  error in  the  determination  of  the ratio  $F_R/F_I$.   In
addition we  should note that  the lower envelope could  not reproduce
accurately  the secondary's  star lightcurve  since there  are several
factors  which  distort  the  ellipsoidal modulation  that  we  cannot
quantify.


To test if  the flaring activity is simultaneous in  the two bands, or
if  one  lags the  other,  we  calculated cross-correlation  functions
between them.   This analysis identifies correlations  and reveals the
mean lag  between the $R$ and  $I$ variability.  We  used the Discrete
Correlation Function  or $DCF$ \citep{edelson88}.  Our  $DCF$, for the
nights when simultaneous observations  last more than 6\,hours (21, 24
and  25 Aug.), is  shown in  figure~\ref{correl}.  Here  positive lags
correspond  to  $R$  variations  lagging  behind those  in  $I$.   The
correlation is clearly significant at the 3$\sigma$ level although the
peak  appears asymmetric.   This  could correspond  to $R$  variations
lasting longer  than those in  $I$.  According to  \citet{tonry79}, we
estimate a  delay of $R$ with  respect to $I$  of (49$\pm$40\,s).  The
measured delay is clearly not significant at the 3\,$\sigma$ level and
is below the time resolution used, so is almost certainly not real. We
hence conclude  that both  the $R$ and  $I$ emission  are simultaneous
within the observational errors.  In fact this is the result we expect
since both bands are close  in wavelength and so probably originate in
the same region.

\subsection{The 2001 simultaneous $R$ and infrared photometry}

Because the  fractional quiescent emission rises  toward the infrared,
it is  of considerable interest  to investigate the properties  of the
flare  emission in  this band.   Even  in the  infrared the  remaining
contamination due to flares can  have significant impact on the binary
parameters  derived  from modeling  the  ellipsoidal variations  (e.g.
\citealt{sanwal95,   shahbaz96}).    We   also   observed   V404   Cyg
simultaneously at optical and infrared wavelengths during 2001.  After
performing  aperture  photometry  around  the  target  and  its  close
companion,   we   subtracted   the   latter  by   assuming   $J$=14.53
\citep{casares93} and  $K$=14.3 \citep{shahbaz96}.  The  lightcurve is
shown in figure~\ref{elipRIR},  where fast variability is superimposed
on the ellipsoidal modulation.\\

We removed the long--term  ellipsoidal modulation by again subtracting
the lower envelope as in section~\ref{curvas_v4R}. The non-ellipsoidal
variability is  shown in figure  ~\ref{curvaRIR}.  To obtain  the same
temporal resolution in  all filters, we first binned  the $R$ data and
then performed linear interpolation of the $R$ and $K_s$ magnitudes to
the  times when  the $J$  magnitudes are  given.  We  note variability
above the noise level, also in the infrared bands.  In some nights the
infrared emission is clearly correlated with the optical (7, 8, 10 and
28    Aug.),    with    amplitudes    $\Delta{J}\sim$0.16\,mags    and
$\Delta{K_s}\sim$0.08\,mags.  On  the night of 3 Aug.   the $K_s$ flux
shows  a steep  variation with  0.1  mags amplitude  and $\sim$1  hour
duration not correlated with the optical or $J$ band flux.\\

In this  case we also obtained the  flux densities as was  made in the
previous section  using the  extinction law by  \cite{rieke85}.  These
fluxes   in    the   regions   with    simultaneous   photometry   are
$F_R$=0.97\,mJy,   $F_J$=0.68\,mJy   and  $F_K$=0.23\,mJy.    Assuming
$F_R/F_I\sim3$ and a $\sim$80\,percent  uncertainty in this ratio (see
previous section) we estimated the flux density in the $I$ band.  Also
we compared  our values with  those obtained by  \citet{shahbaz03b} to
extract the spectrum of the flares (Fig~\ref{spec}). These last values
are  actually  the fluxes  at  the peak  of  individual  flares so  we
consider them as  upper limits. Apart from this, we  can only obtain a
very  rough estimation  since  we are  comparing  data from  different
epochs and with large uncertainties. \\

In  order  to  characterize  the  flare  activity,  we  calculate  the
parameters of  the variability in  the three bands, the  results being
given in  table~\ref{flares_RIR}.  We  found lower activity  at longer
wavelengths  although, as  we will  discuss in  the next  section, the
relative contribution  of the accretion  disc depends on  the waveband
and hence a further correction must to be made.\\

\begin{table*}
 \centering
\begin{minipage}{140mm}
  \caption{Flares properties for optical and infrared in 2001.}
  \label{flares_RIR}
  \begin{tabular}{@{}lccccc@{}}
\hline
{\em  Band}  &  {\em  $<\Pi^{\dagger}>$} &  {\em  Monitoring}  &  {\em
$<L>$} &  {\em $\bar{z_f}$} & {\em $\sigma_z$}  \\ & & {\em  time (hr)} &
{\em $\times$10$^{32}$\,erg\,s$^{-1}$} & & \\
  \hline
R & 0.071 & 21.1 & 13.5$\pm$0.4 & 0.069 & 0.036\\
J & 0.057 & 21.1 &  7.7$\pm$0.3 & 0.058 & 0.029\\
K & 0.035 & 21.1 &  1.8$\pm$0.1 & 0.029 & 0.012\\
\hline
\end{tabular}

\medskip
($\dagger$) Time averaged equivalent duration of the individual
flares divided by the total monitoring time.
\end{minipage}
\end{table*}


\section{Discussion}

Our $R$  band lightcurves, obtained during the  period 1992-2003, show
considerable  suborbital variability  in addition  to  the ellipsoidal
modulation.   Assuming that  these  variations are  not connected,  we
obtained a representation of the ellipsoidal modulation of the tidally
distorted  secondary  star by  fitting  a  double  sine to  the  lower
envelope of the lightcurve. The  fits performed with the most complete
databases (years 1992,  1998 and 2001) show unequal  maxima and minima
with relative strength varying from year to year.  This behavior could
be explained assuming starspots on  the secondary star. This is a very
widespread phenomenon  among cool stars.  Axial rotation  of a spotted
star and  slow variations of  the starspot geometry  cause photometric
variability which is expressed in brightness rotational modulation and
slow variations of mean stellar lightcurves. Other possibility is they
are produced by changes in  the accretion disc geometry and brightness
due  a superhump,  i.e.  optical  modulation as  a consequence  of the
precession  of an eccentric  accretion disc  by perturbation  from the
secondary. This is  a very promising suggestion since  superhumps in a
quiescent    SXT    have    already    been    detected    (J1118+480,
\citealt{zurita02}).   Superhumps have  also been  largely  invoked to
explain the  changing quiescent  lightcurve of the  prototypical black
hole candidate: A0620-00 (eg. \citealt{haswell96}).  After subtracting
the  corresponding  lower evelope  fits  from  the  original data  the
lightcurves are dominated by  large flares ($\sim0^m.2$) lasting a few
hours.  In  addition, shorter timescale  variability ($\sim0^m.05$) is
present  with  lower amplitude.   Our  photometry  indicates that  the
flaring  activity  did  not  change significantly  over  this  10-year
interval  which  indicates  that  this  variability  is  probably  not
connected to the 1989 outburst.\\

We have also found significant variability in the $I$ and near-IR band
($J$  and $K_s$) and  these are  correlated.  In  1990, just  one year
after the  outburst, a  variation of $\sim0^m.2$  was seen  in several
visible bands \citep{casares93}.   In 1991, the variability originally
interpreted as a six hour modulation, decreased to $\sim0^m.1$, but no
significant   changes  were   reported  in   the   interval  1992-1993
\citep{martin99}.   Infrared photometry  in 1990,  was dominated  by a
$\sim0^m.2$   amplitude    at   a   periodicity    of   $\sim$6\,hours
\citep{casares93}.   Also  1993  $H$  band data  presents  significant
variability contaminating the ellipsoidal modulation \citep{sanwal95}.
They  reported an  $rms$  scatter  of $\sim0^m.04$  for  V404 Cyg  and
$\sim0^m.02$ for  the comparison star.   This yields a  flare activity
level of $\sigma_z\sim$0.032, which is consistent with our values.  In
1993, $K$ band photometry  obtained by \citet{shahbaz94} does not show
short term variability, with  an upper limit of $\sim0^m.03$.  However
in  this case,  only one  night of  more than  6\,hours  of continuous
coverage was obtained. \\

We have also found that lower activity is found at longer wavelengths.
However,  as we  have  pointed out  in  ZCS the  flare activity  level
calculated as $\sigma_{\rm z}$  represents the flare activity relative
to  the  underlying smooth  modulation  ($F_q$).   $F_q$ contains  two
distinct components: the  flux from the companion star,  $F_c$ and the
flux from the non-variable, steady accretion disc, $F_d$ (see figure 9
in  $ZCS$ for clarity).   The relative  contribution of  the accretion
disc depends on the wave band  and hence, if we assume that the flares
are  produced in the  accretion disc  (see \citealt{hynes02},  ZCS), a
more useful parameter to compute is the flare activity relative to the
flux  from  the  steady  disc,  rather  than  relative  to  the  total
non-variable flux, which is given by

\begin{equation}
\sigma_{\rm z}^{*} = \sigma_{\rm z} / \eta_{\rm d} 
\end{equation}

\noindent
where 

\begin{equation}
\eta_{\rm d} = F_{\rm d} / F_{\rm q} = (1 + F_{\rm c}/F_{\rm d})^{-1}
\end{equation}

We can determine $\eta_{\rm d}$ using the spectroscopic value for the veiling
$\eta_{\rm obs}$, which is defined as

\begin{equation}
\eta_{\rm obs} = \frac{\bar{F_{\rm f}}+F_{\rm d}}{\bar{F_{\rm f}}+F_{\rm q} }
\end{equation}

\noindent
where $\bar{F_{\rm f}}$ is the mean flux of the flares, and find

\begin{equation}
\label{vd_spec}
\eta_{\rm d}=(\bar{z_{\rm f}}+1)\,\eta_{\rm obs}-\bar{z_{\rm f}}
\end{equation}

\begin{table*}
 \centering
\begin{minipage}{140mm}
  \caption{Flare activity and veiling factors for $R$, $I$ and $K$ band.}
  \label{veiling}
  \begin{tabular}{@{}lccccc@{}}
\hline
{\em Band}  & {\em  $\eta_{\rm obs}$(percent)} &  {\em $\eta_{\rm d}$(percent)}  & {\em
$\bar{z_f}$} & {\em $\sigma_z$} & {\em $\sigma_z^{*}$} \\
  \hline
R & 13 & 14.1 & 0.095 & 0.030 & 2.1$\times$10$^{-3}$\\
\vspace{3mm}
I & 12 & 12.4 & 0.039 & 0.019 & 1.5$\times$10$^{-3}$\\

R & 13 & 13.8 & 0.069 & 0.036 & 2.6$\times$10$^{-3}$\\
K &  5 &  5.1 & 0.029 & 0.012 & 2.3$\times$10$^{-3}$\\
\hline
\end{tabular}
\end{minipage}
\end{table*}

\citet{casares93} reported  that the accretion  disc contributes 11-16
 percent    of    the    observed    flux    in    $R$    (see    also
 \citet{casares94}). Therefore  we assume an  observational veiling of
 $\eta_{\rm   obs}=13\%$  in   this   band.   We   also  estimate   an
 observational veiling of $\sim12\%$  in the $I$ band by extrapolating
 from the veiling in $B$, $V$ and $R$ \citep{casares93}.  On the other
 hand, \citet{shahbaz96} determined the  fraction of the light arising
 from the  accretion disc in  the $K$ band  varies from 0 to  an upper
 limit of 11 percent.  We hence consider $\eta_{\rm obs}$=5 percent in
 this case.   The corrected values using these  factors, are tabulated
 in  table~\ref{veiling}. The  uncertainties  on the  veiling and  the
 limited database only  allows us to say that  the activity diminishes
 slowly or remains approximately constant through the IR.  In the 1992
 series of  $B$, $V$, $R$  and $I$ band observations  (16 measurements
 spread  over 16 nights;  see \citealt{martin99})  an estimate  of the
 variability  could be  obtained by  comparing the  variance  with the
 photometric noise and the optical variations. This variance increases
 toward the  blue suggesting  that the source  of fast  variability in
 V404  is  indeed  blue.   We  note  however  that  \citet{shahbaz03b}
 determined  the  colour  of  the  large flares  and  found  the  $i'$
 ($\lambda_{eff}7650$\AA) band flux to be larger than that in the $g'$
 ($\lambda_{eff}4750$\AA) band. Comparing  with these data we obtained
 an  estimated spectrum  with  a peak  at $\lambda_{eff}7000$\AA  (see
 Fig.~\ref{spec}).\\

In an attempt to compare the infrared variability we also observed the
SXT with  highest flares:  J0422+32, where events  as high as  0.6 mag
where observed  (ZCS).  Observations  of this system  in the  $H$ band
covering  half of  its orbital  period do  not show  variability above
$\sim$0.06\,mags,  the estimated  error of  our photometry  (Zurita et
al. in  preparation).  Comparing with  the variability we  observed in
the $R'$ band (ZCS) and  assuming that the disc contributes $\sim$30\%
in $H$  (Beekman et  al. 1997)  we estimate that  the activity  in the
infrared is, at least, 33\% lower than in the optical. Although only a
very  crude estimation  can be  made, among  other things  because the
observations are not simultaneous and because the value of the veiling
is not well determined, it  seems that the spectrum of J0422+32 flares
also drop  at longer wavelengths.   On the other hand,  no significant
contamination of  the infrared lightcurves of J0422+32  has been found
by \citet{gelino03}.\\

The quasi periodic  oscillations noticed in the earliest  data (1992 -
see \citealt{pavlenko96}) are less clear in our more recent data (1998
and 2001).  However, the periodograms show peaks close to 6\,hr at the
68  percent  significance  level  which  could  be  interpreted  as  a
consequence of  the appearance  of a flare  event almost  every night.
\citet{shahbaz03} found a $QPO$ feature at 21.5\,minutes attributed to
variability at the $ADAF$  transition radius. Our longer sampling does
not allow us to test this  hypothesis. It is possible in any case that
variability with different timescales have different origin.\\


\section*{Acknowledgments}

CZ  acknowledges support  from the  Funda\cao\  para a  Ci\^encia e  a
Tecnologia, Portugal.   IH acknowledges support  from grant F/00-180/A
from  the  Leverhulme  Trust   and  through  Hubble  Fellowship  grant
\#HF-01150.01-A  awarded  by the  Space  Telescope Science  Institute,
which is operated  by the Association of Universities  for Research in
Astronomy, Inc., for NASA, under contract NAS 5-26555. TS acknowledges
support  from the  Spanish Ministry  of Science  and  Technology under
project AYA 2002 03570.

\newpage

\label{parametros_v4}
\begin{figure}
\includegraphics[width=8.cm]{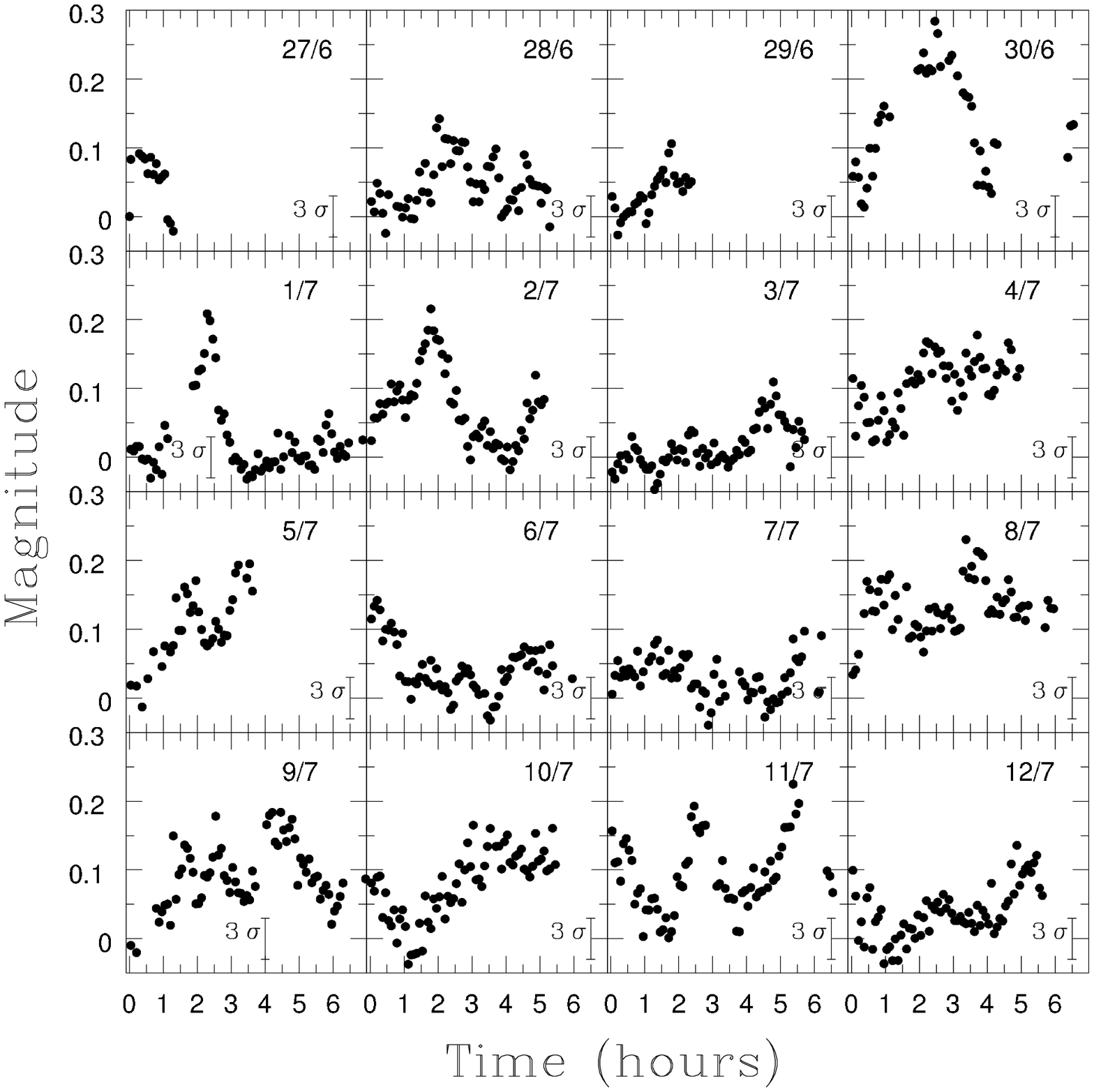}
\caption{$R$ band lightcurves of V404 Cyg after subtracting the ellipsoidal 
modulation -- 1992 lightcurves.}
\end{figure}
\addtocounter{figure}{-1}
\begin{figure}
\includegraphics[width=8.cm]{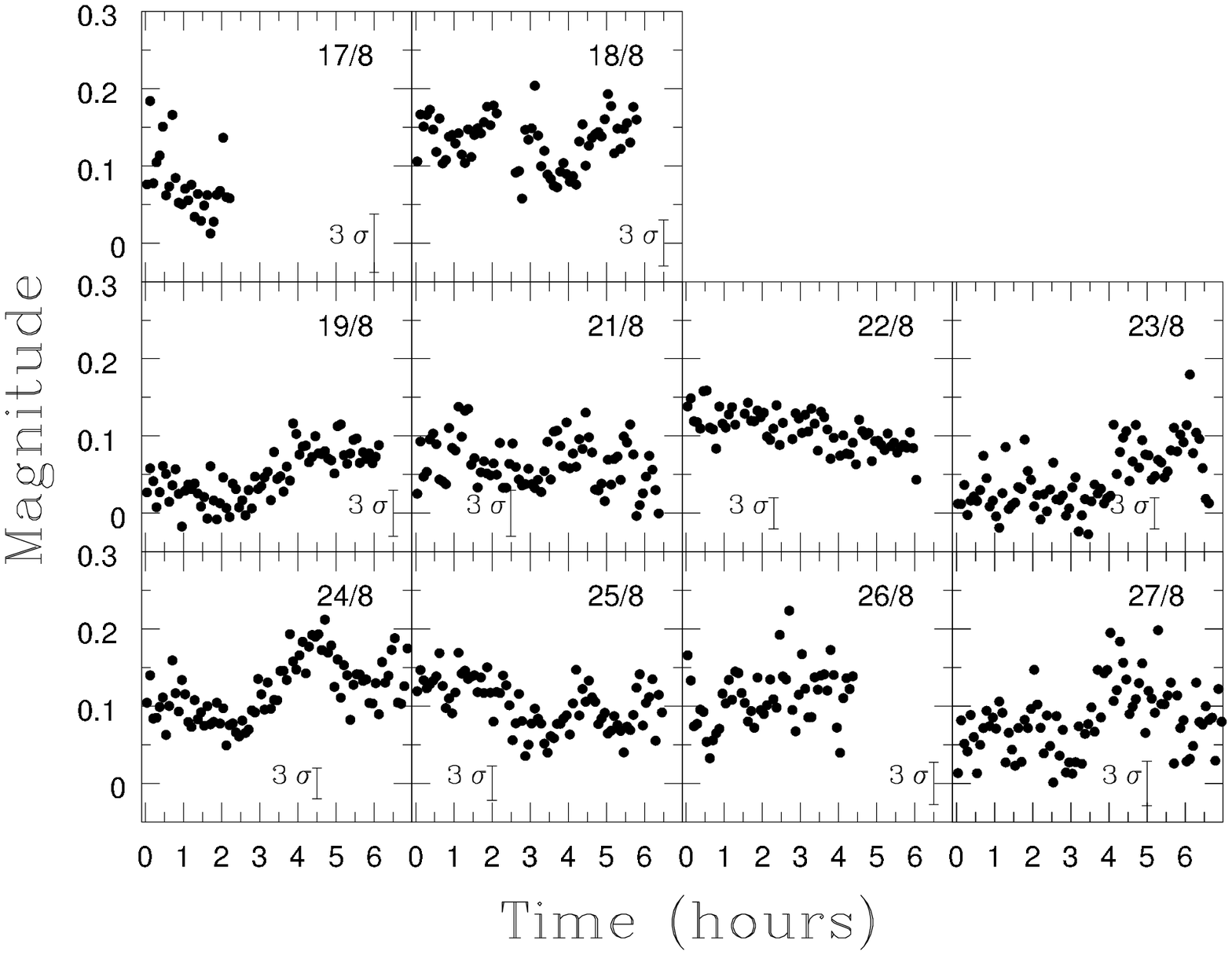}
\caption{$R$ band lightcurves of V404 Cyg after subtracting the ellipsoidal 
modulation -- 1998 lightcuurves.}
\end{figure}
\addtocounter{figure}{-1}
\begin{figure}
\includegraphics[width=6.cm]{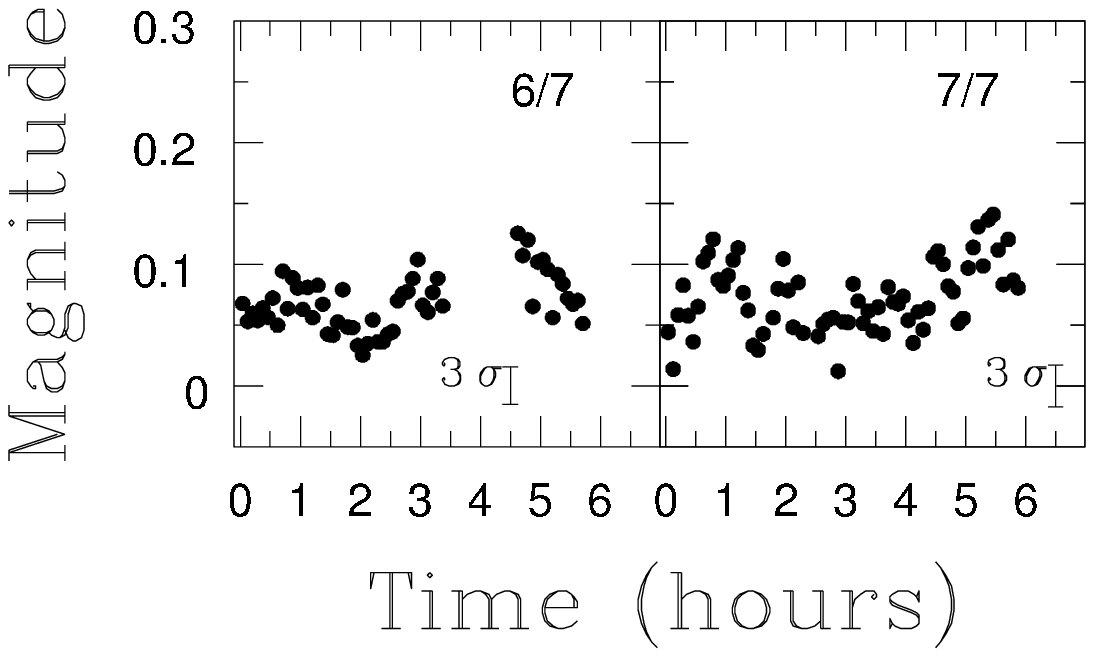}
\caption{$R$ band lightcurves of V404 Cyg after subtracting the ellipsoidal 
modulation -- 1999 lightcurves.}
\end{figure}
\addtocounter{figure}{-1}
\begin{figure}
\includegraphics[width=8cm]{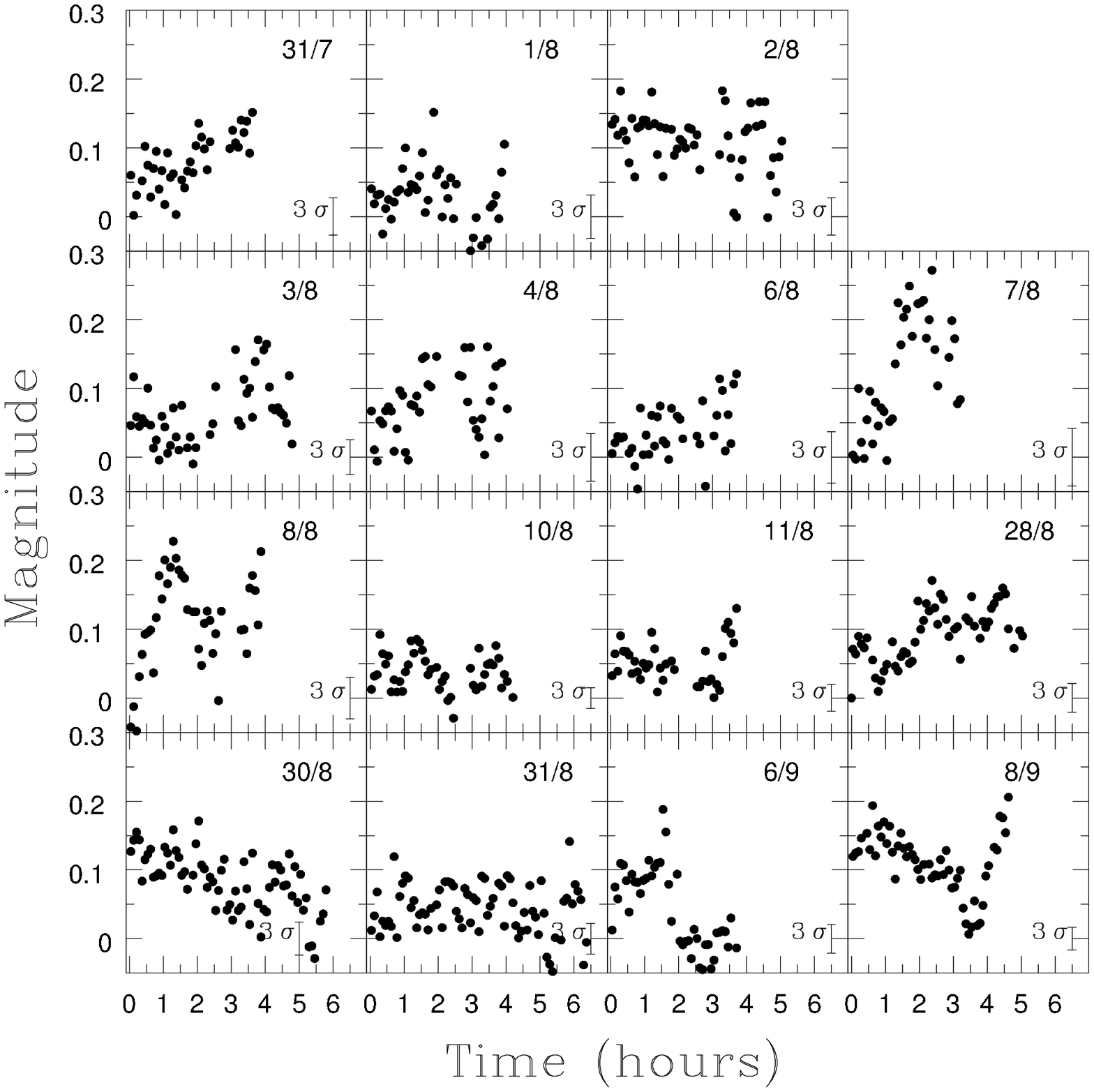}
\caption{$R$ band lightcurves of V404 Cyg after subtracting the ellipsoidal 
modulation -- 2001 lightcurves.}
\end{figure}
\addtocounter{figure}{-1}
\begin{figure}
\includegraphics[width=8cm]{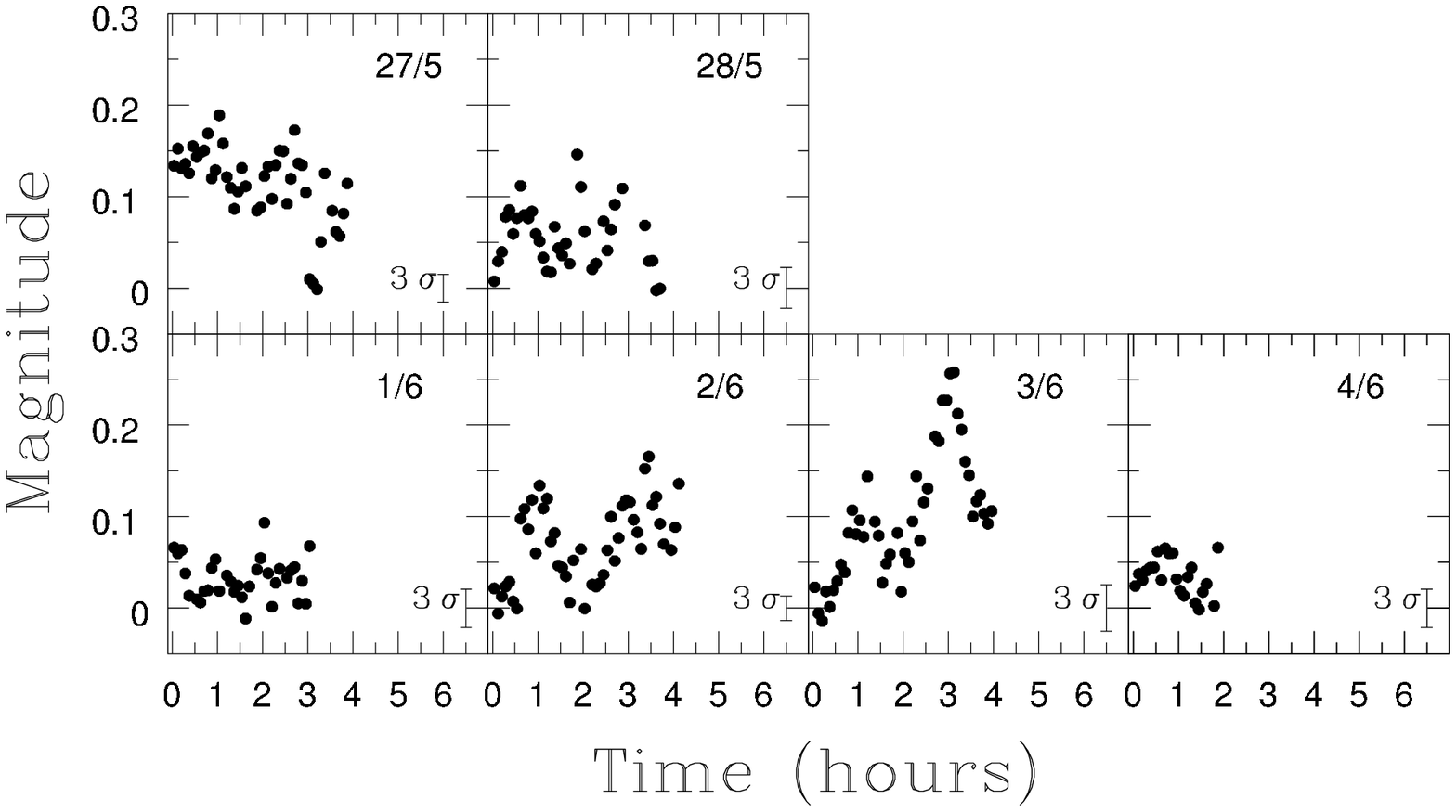}
\protect\caption[Detrended lightcurves]{Continued -- 2002 lightcurves.}
\end{figure}
\addtocounter{figure}{-1}
\begin{figure}
\includegraphics[width=8cm]{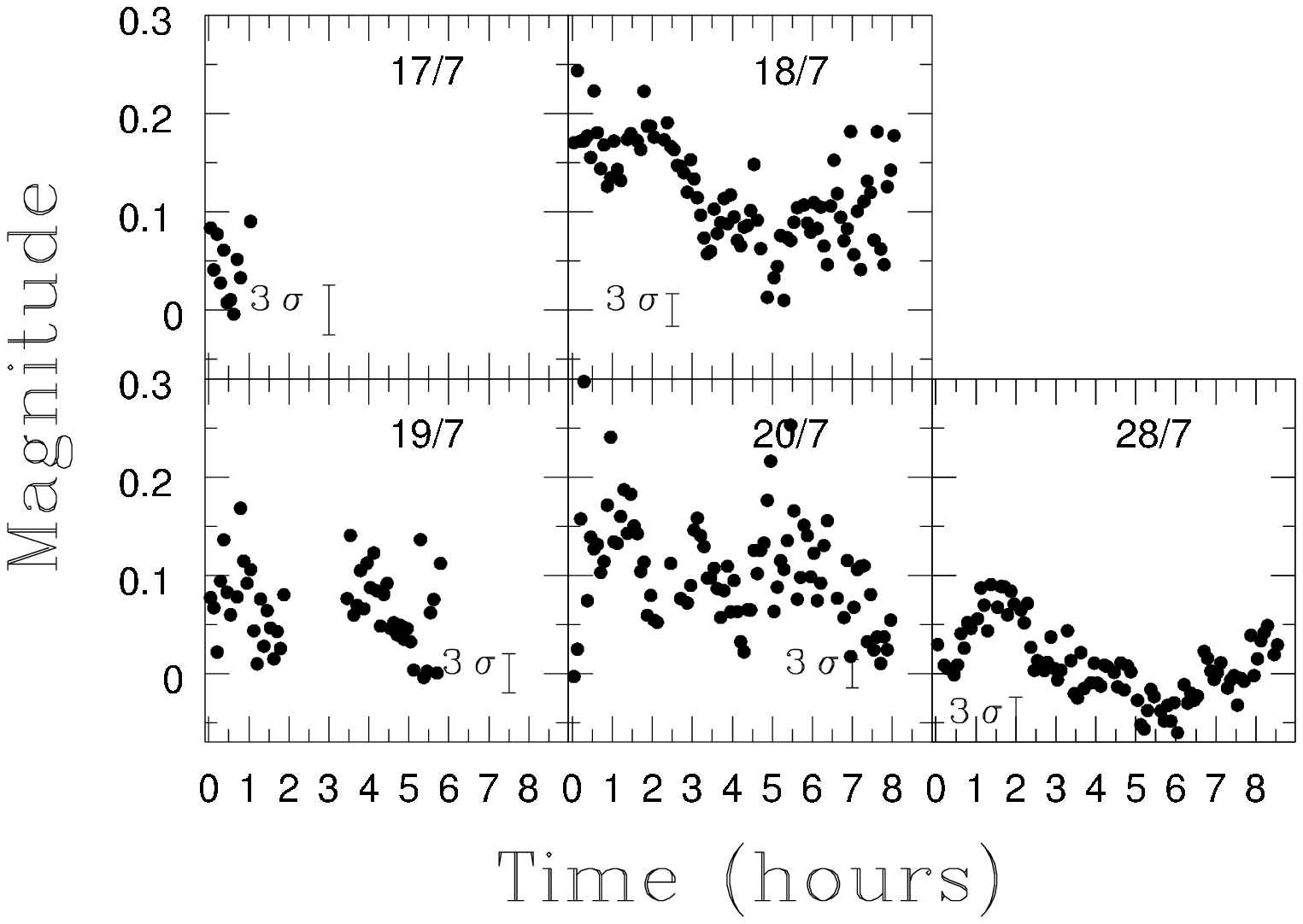}
\protect\caption[Detrended lightcurves]{Continued -- 2003 lightcurves.}
\label{var}
\end{figure}
\begin{figure}
\includegraphics[width=8cm]{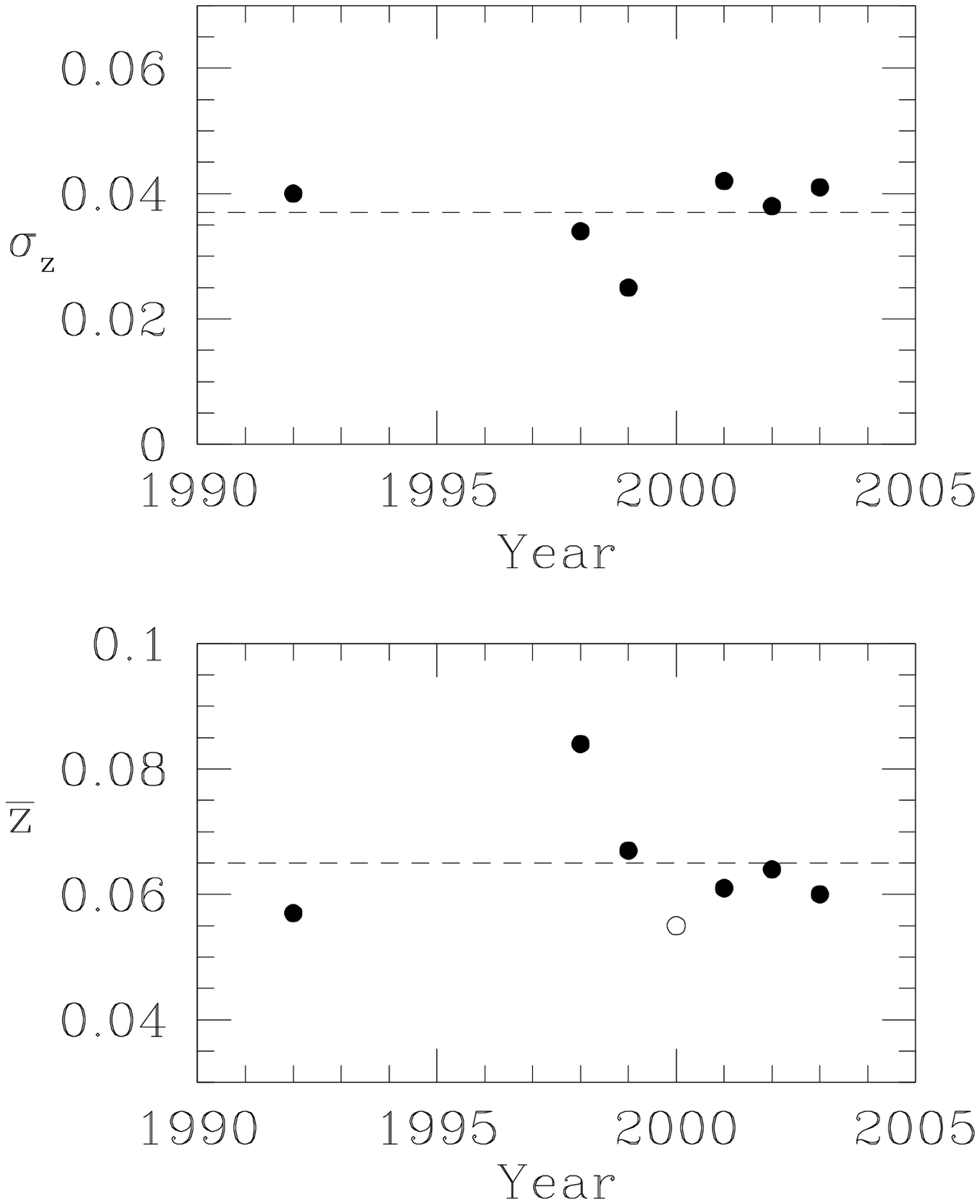}
\caption{Mean  relative intensity  and  flare activity  over the  last
decade.   Mean levels  are marked  by dashed  lines.  The  open circle
marks the value measured by \citet{pavlenko01}.}
\label{sample}
\end{figure}
\begin{figure}
\includegraphics[width=8cm]{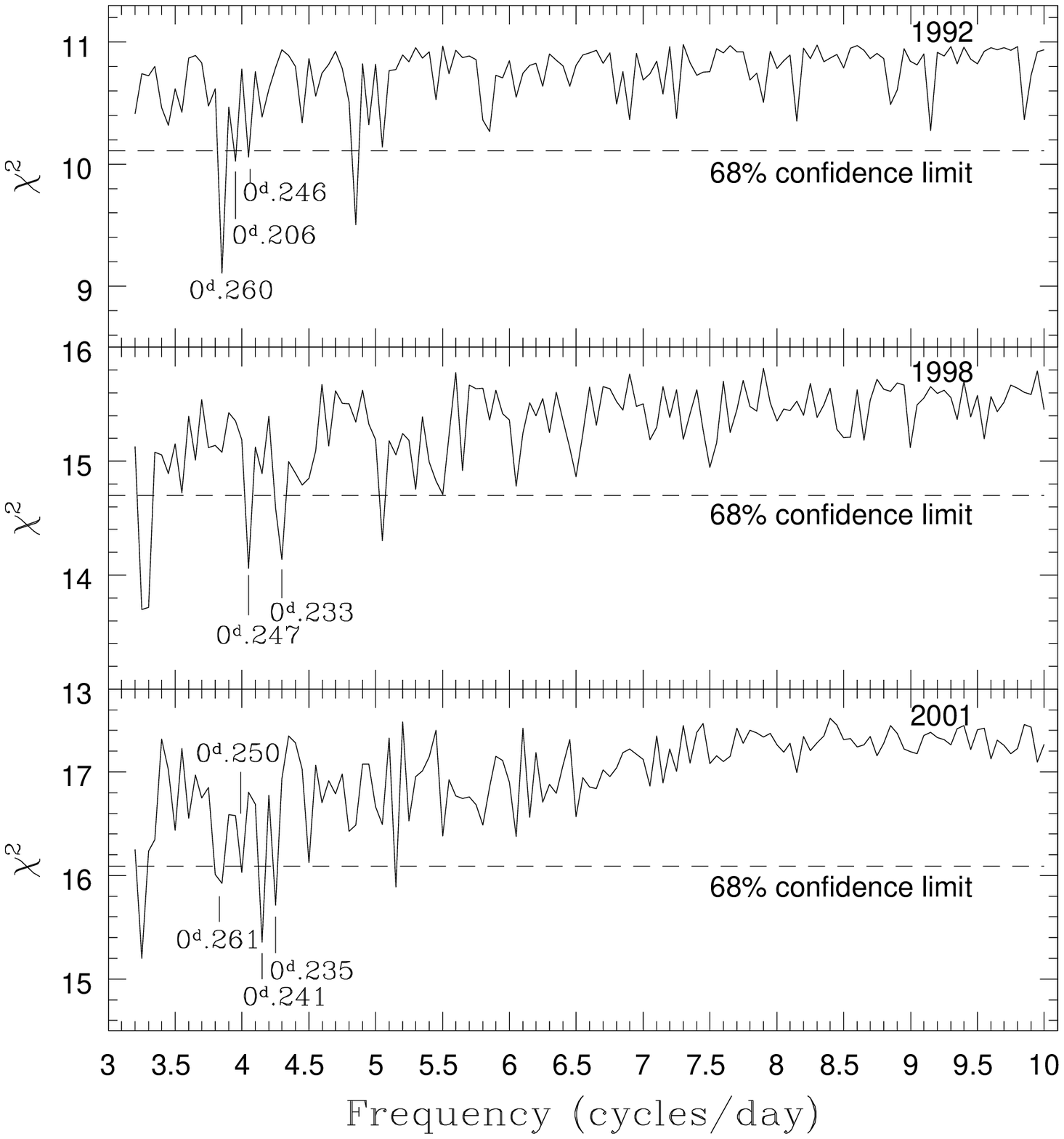}
\caption{$\chi^2$ minimization
spectrum of  the data after subtraction  of the lower  envelope. Dashed lines
marks the 68 percent confidence  limits at $\chi^2_{min}+1$.}
\label{period}
\end{figure}
\begin{figure}
\includegraphics[width=7cm]{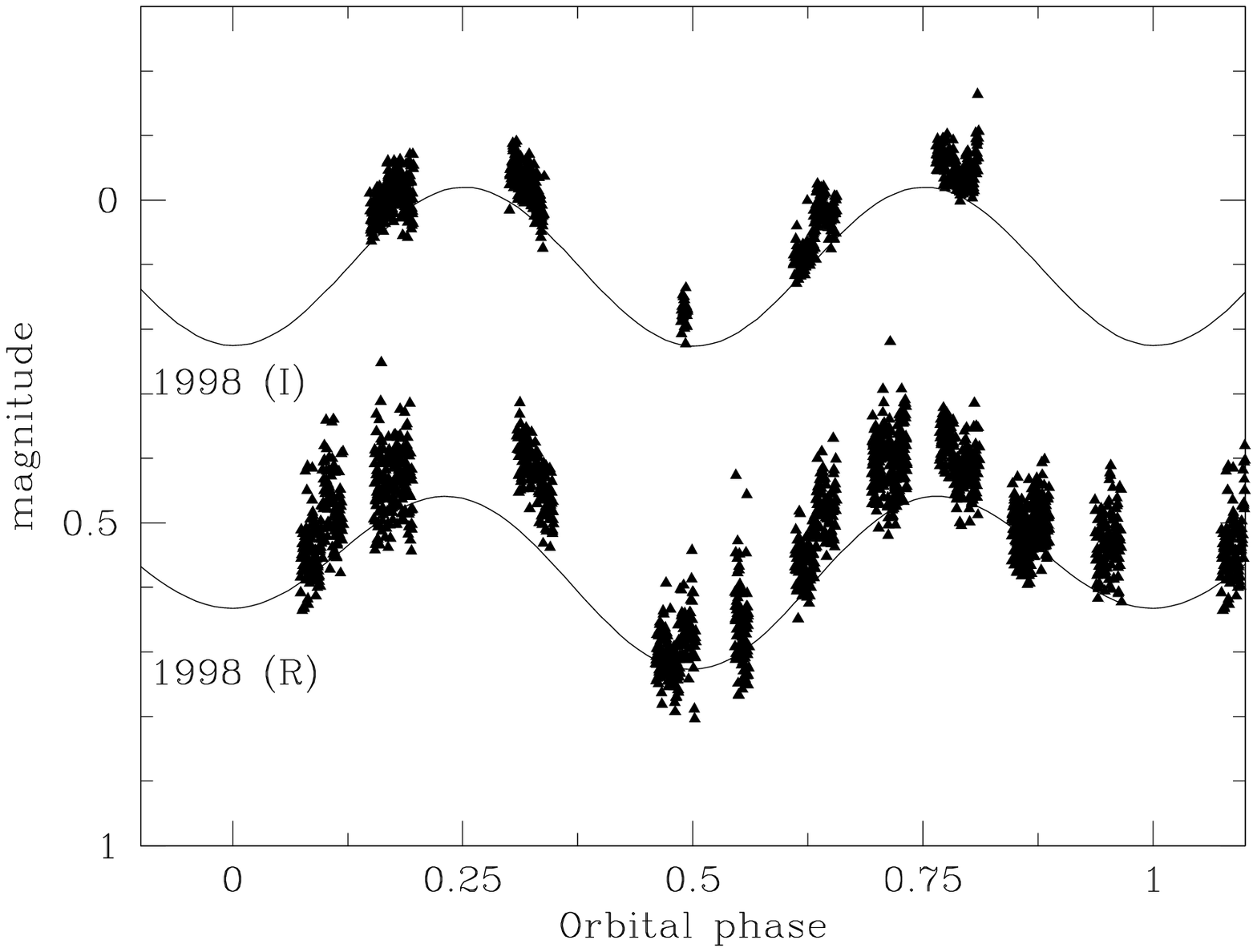}
\caption{The  $R$ and  $I$ 1998  lightcurves and  the  lower envelopes
constructed as explained in section \ref{curvas_v4R}.}
\label{elipI}
\end{figure} 
\begin{figure}
\includegraphics[width=8cm]{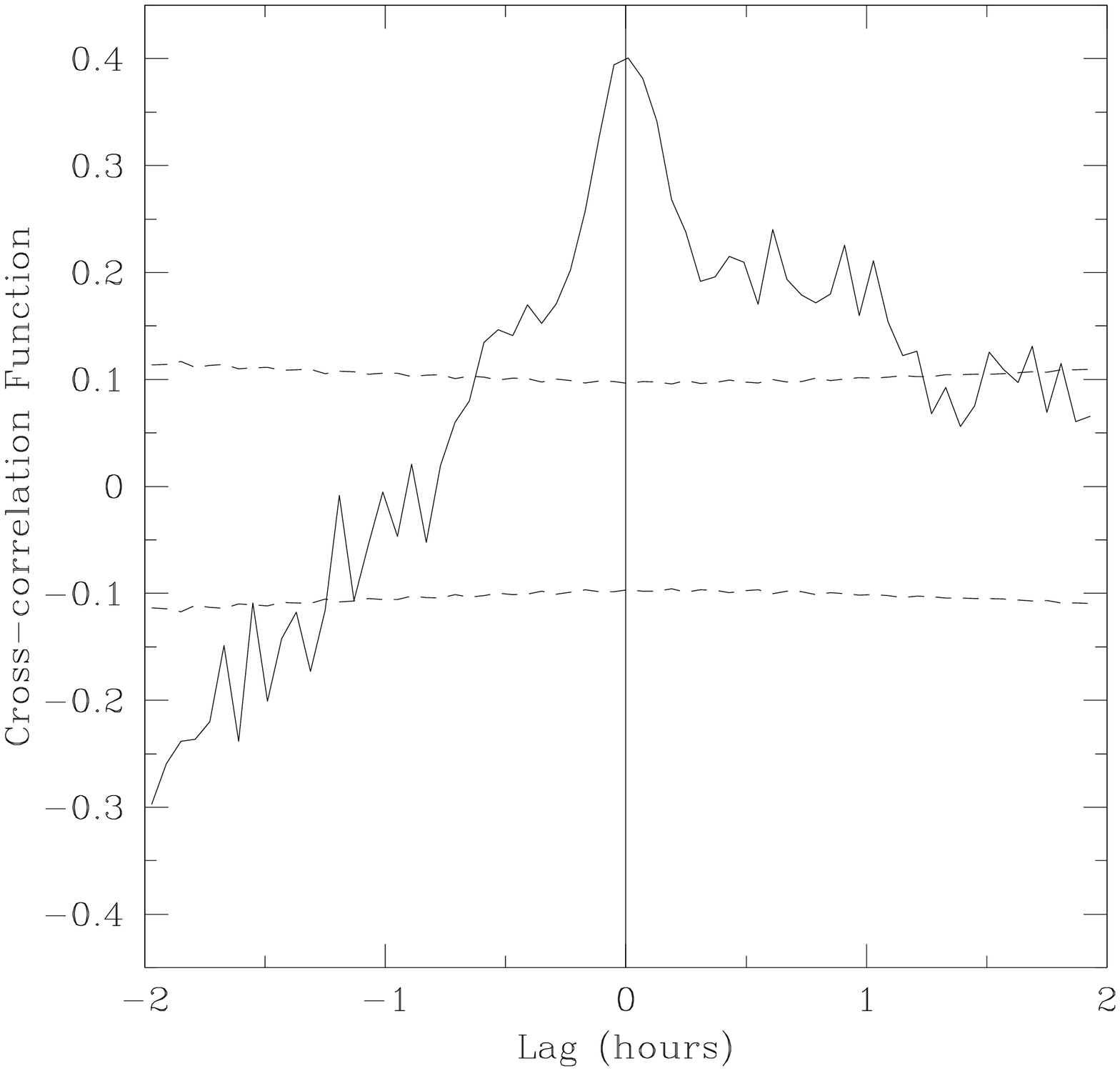}
\caption{Cross-correlation function between $R$  and $I$ band data.  A
positive lag  would correspond to $R$ variations  lagging behind those
in   $I$.  Dotted   lines  indicate   3$\sigma$  limits   on  expected
coincidental correlations.}
\label{correl}
\end{figure} 
\begin{figure}
\includegraphics[width=5.5cm]{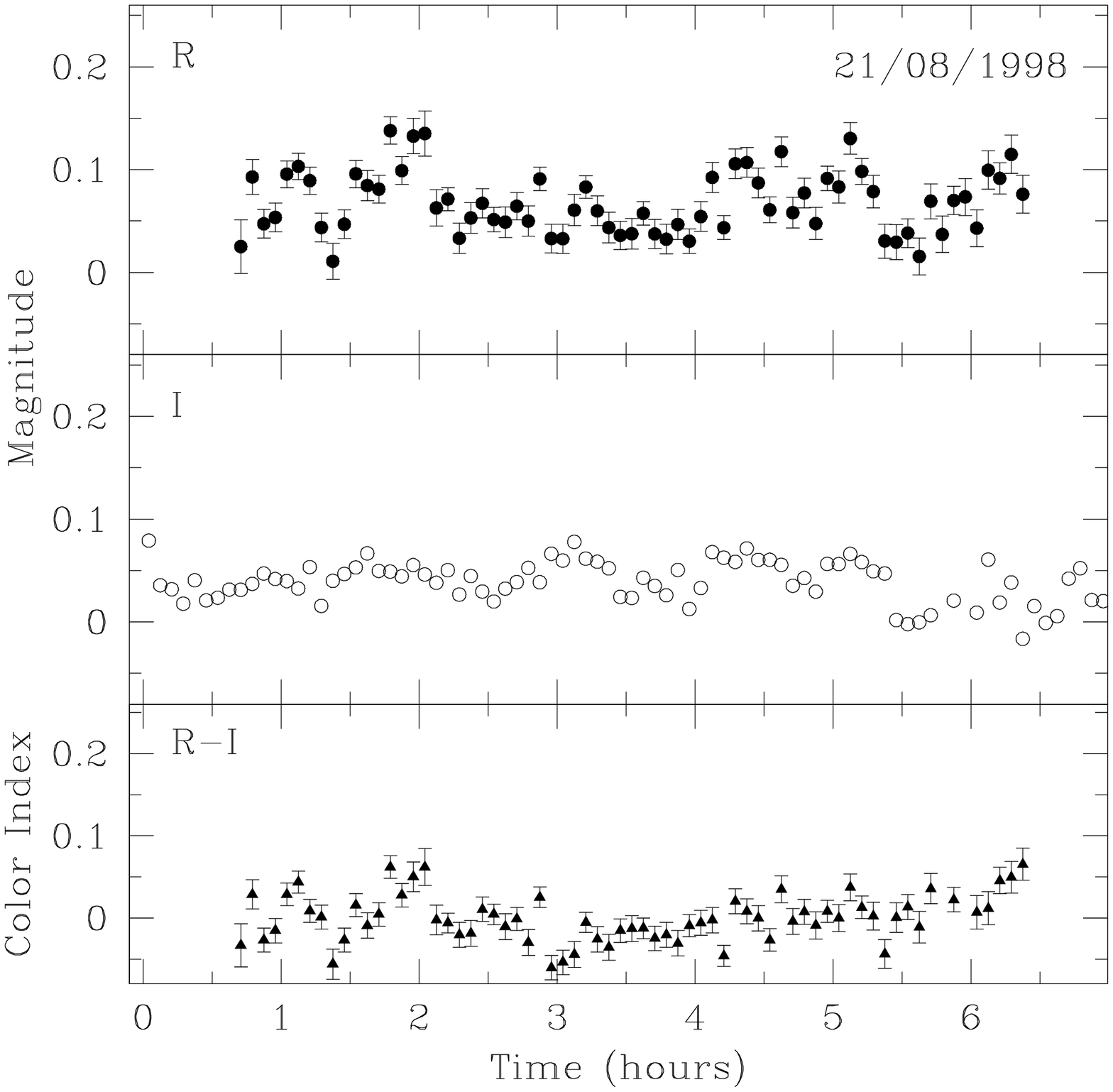}
\includegraphics[width=5.5cm]{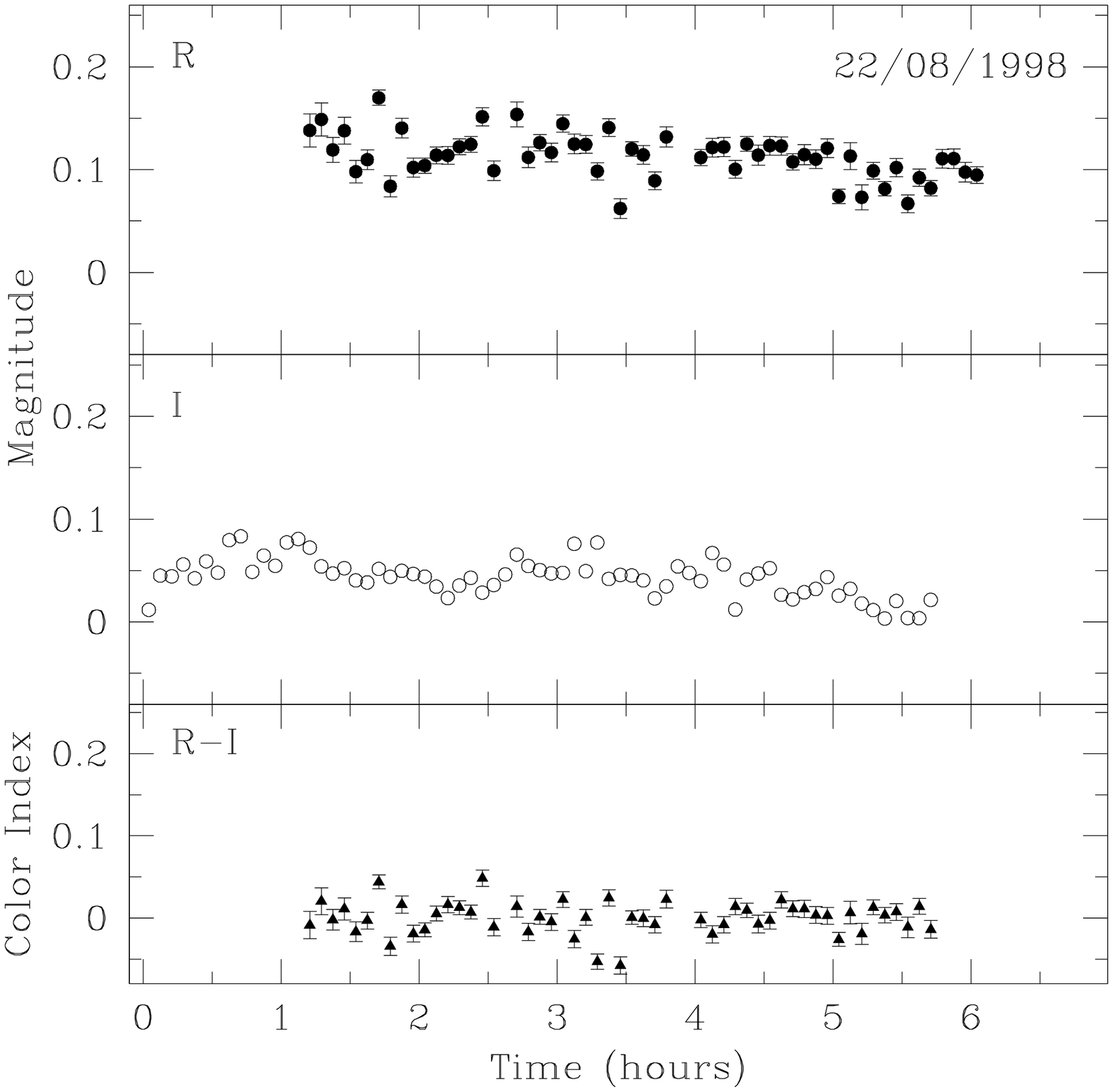}
\includegraphics[width=5.5cm]{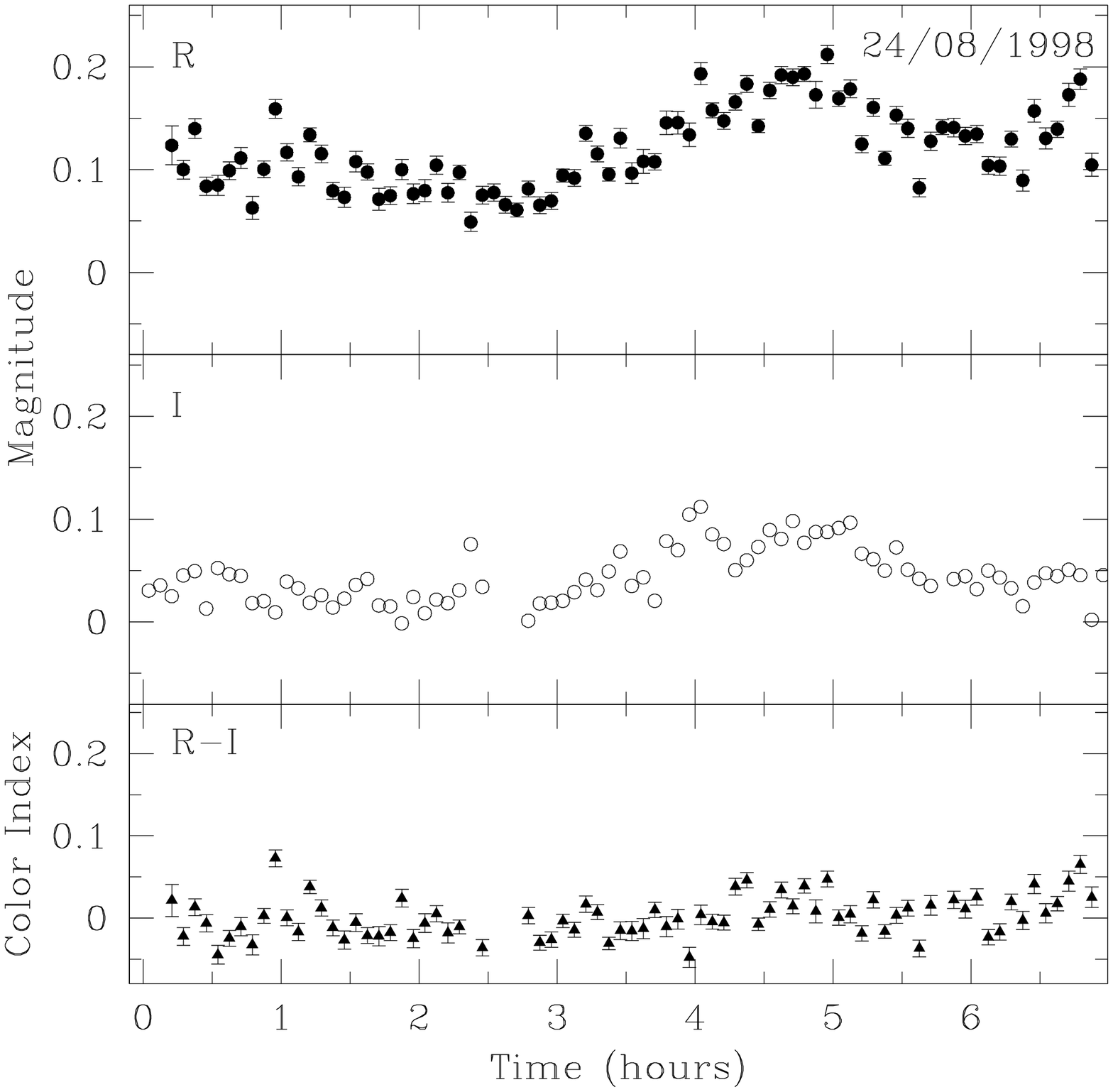}
\includegraphics[width=5.5cm]{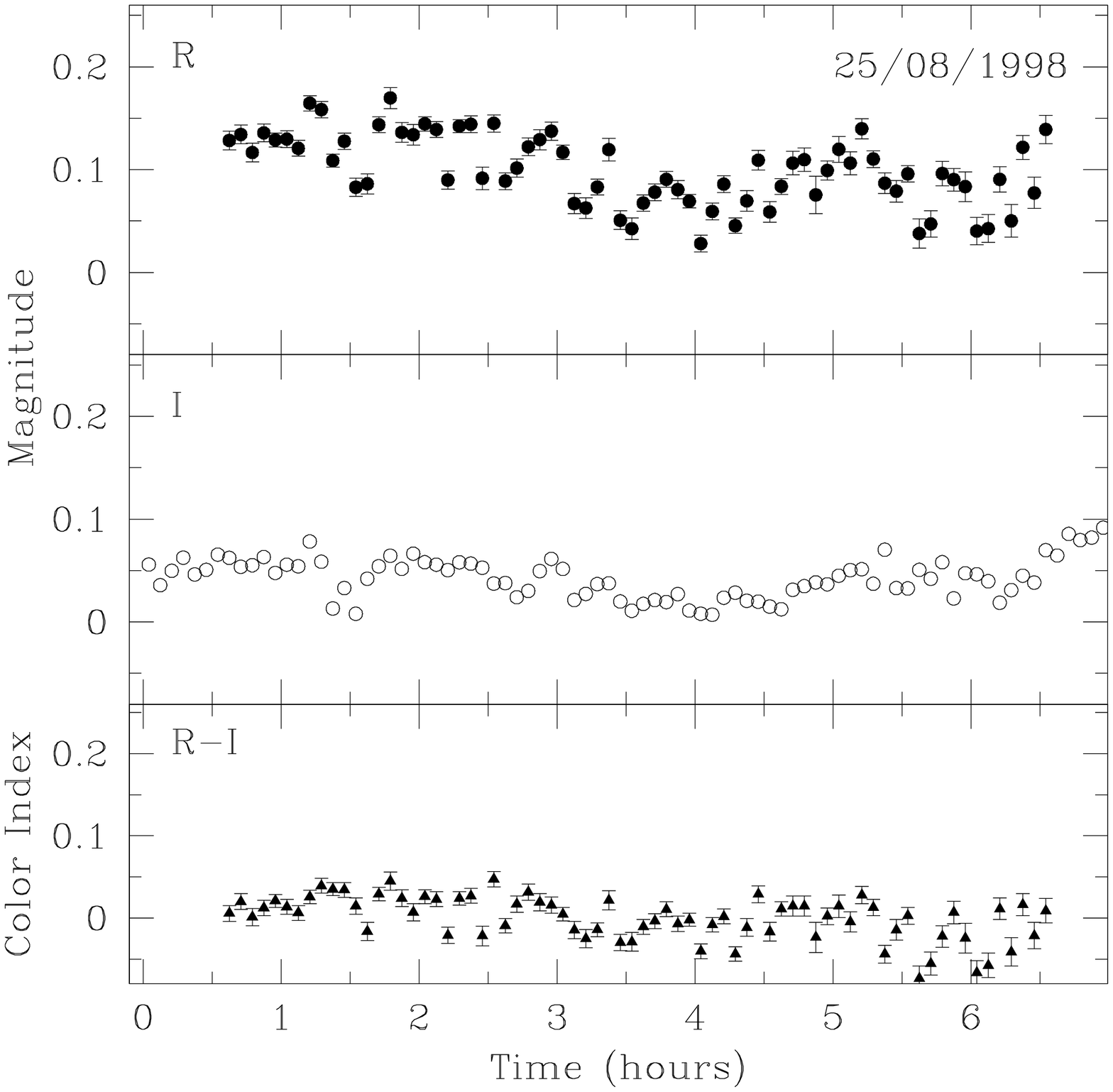}
\caption{The 1998 $R$, $I$ and color $R-I$ lightcurves of V404 Cyg after subtracting the ellipsoidal modulation.}
\label{curvaRI}
\end{figure}
\begin{figure}
\begin{center}
\includegraphics[width=7cm]{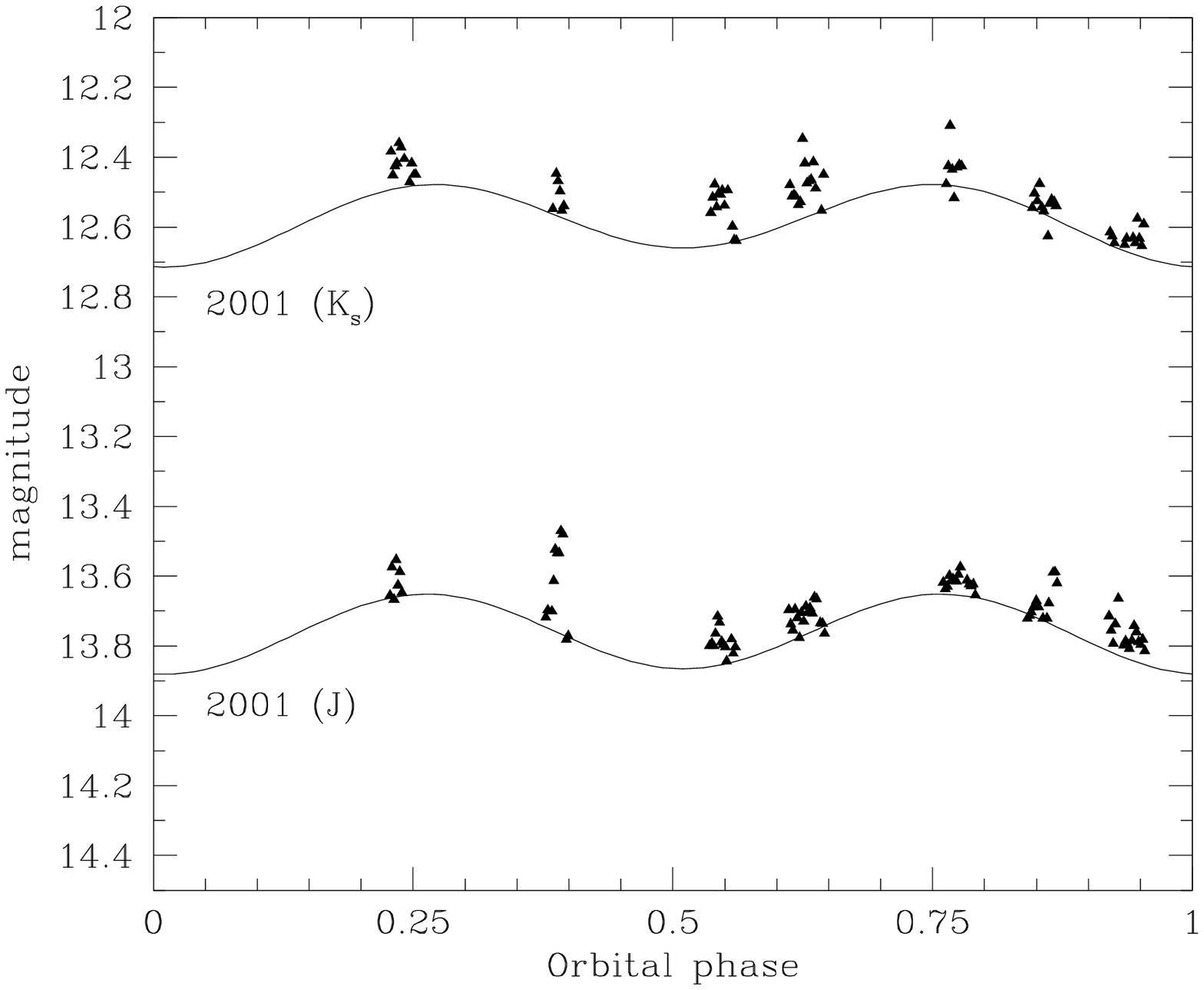}
\caption{The $J$ and $K_s$
ellipsoidal 2001 lightcurves and  the lower envelopes constructed as explained
in section~\ref{curvas_v4R}.}
\label{elipRIR}
\end{center}
\end{figure} 
\begin{figure}
\includegraphics[width=8cm]{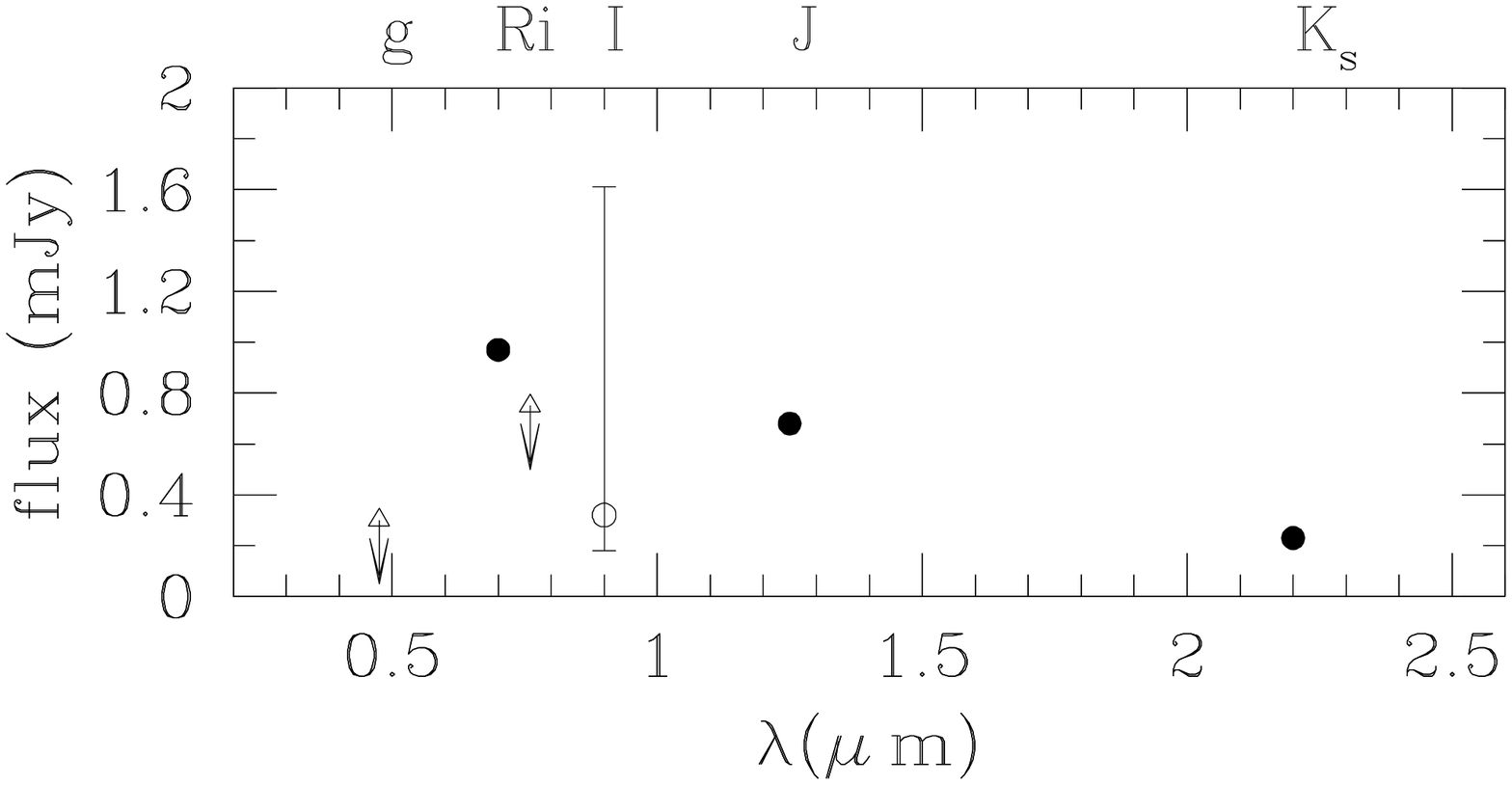}
\caption{Mean flare flux densities in the $R$, $J$ and $K_short$ bands
calculated with the 2001 data  (close circles). The open circle is the
expected $I$ density flux assuming the ratio $F_R/F_I=3$ obtained with
the 1998 data. Triangles are the mean flare flux densities calculated by
\citet{shahbaz03b} in the $g'$  and $i'$ bands ($\lambda_{eff}$= 4750\,
\AA and 7650\,\AA respectively).}
\label{spec} 
\end{figure}
\onecolumn
\begin{figure}
\begin{center}
\includegraphics[width=5cm]{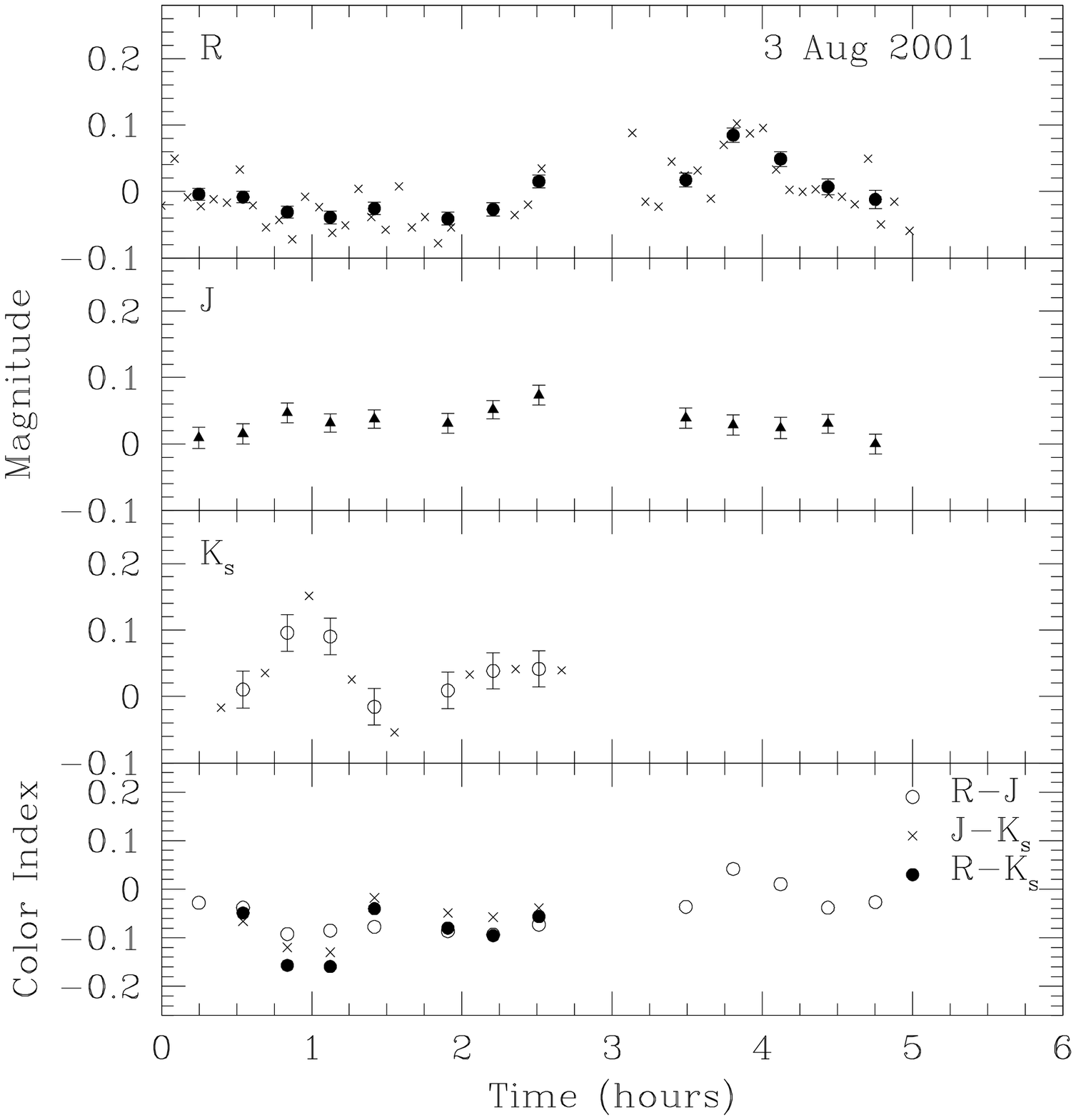}
\includegraphics[width=5cm]{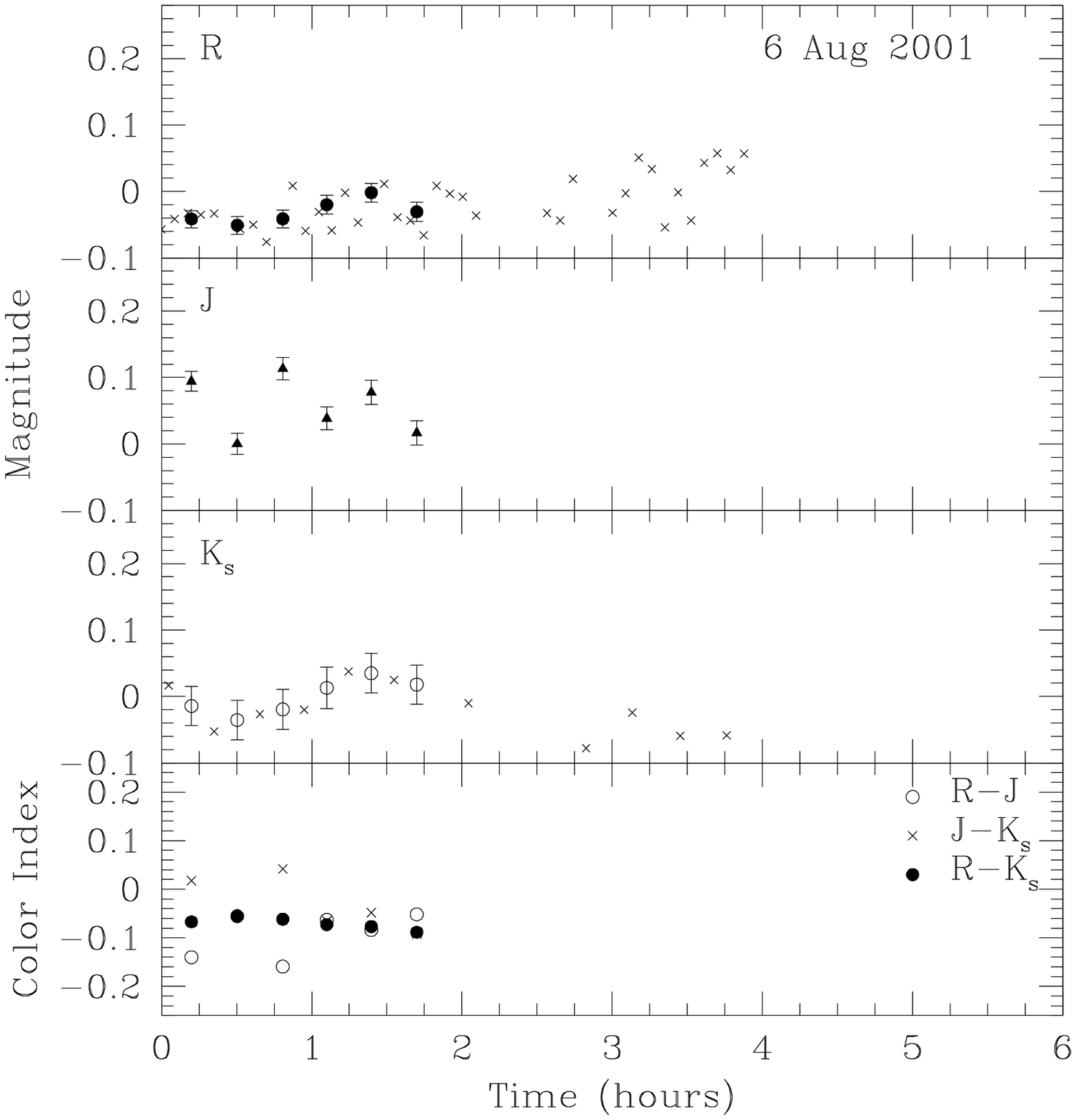}
\includegraphics[width=5cm]{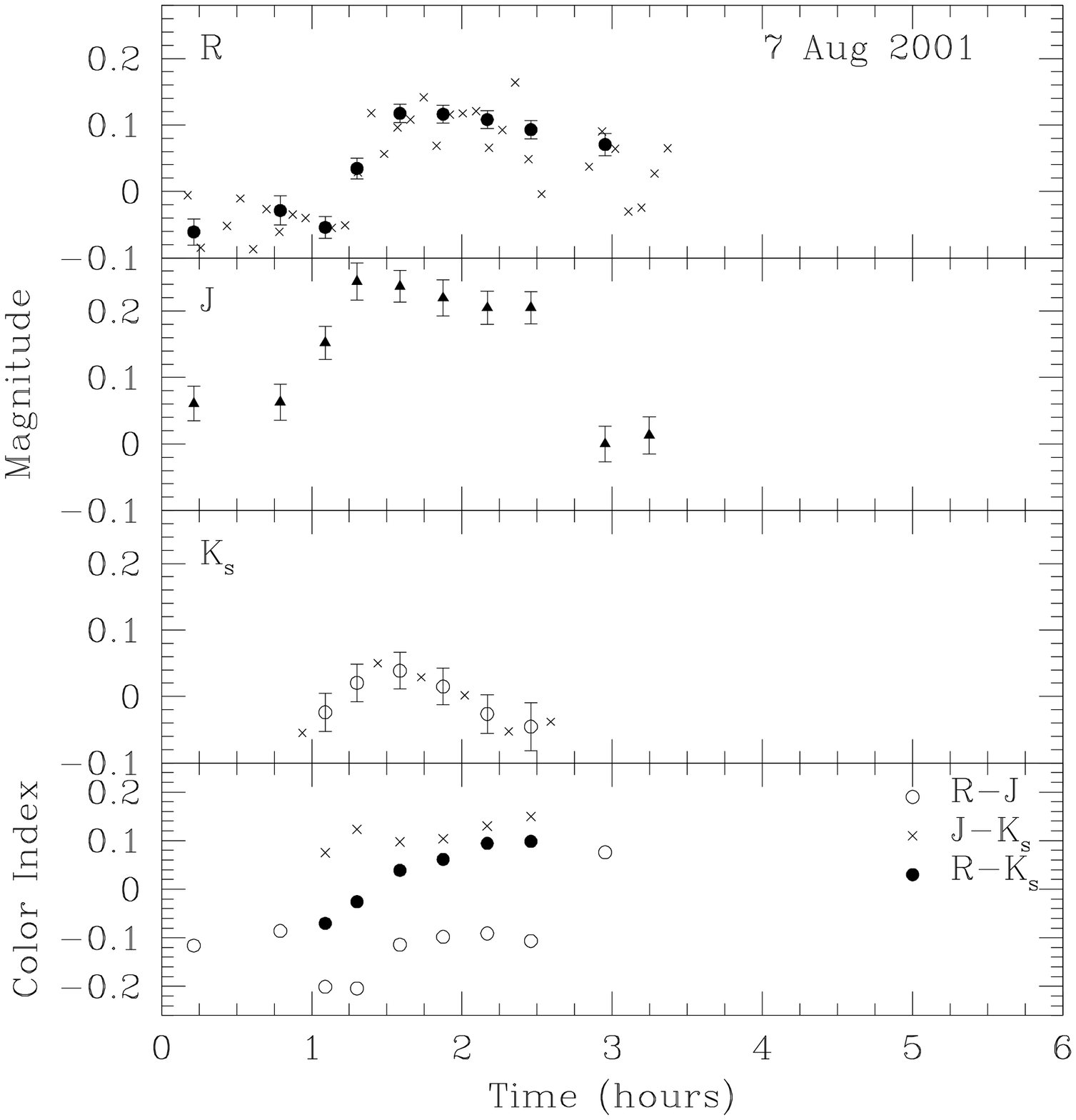}
\includegraphics[width=5cm]{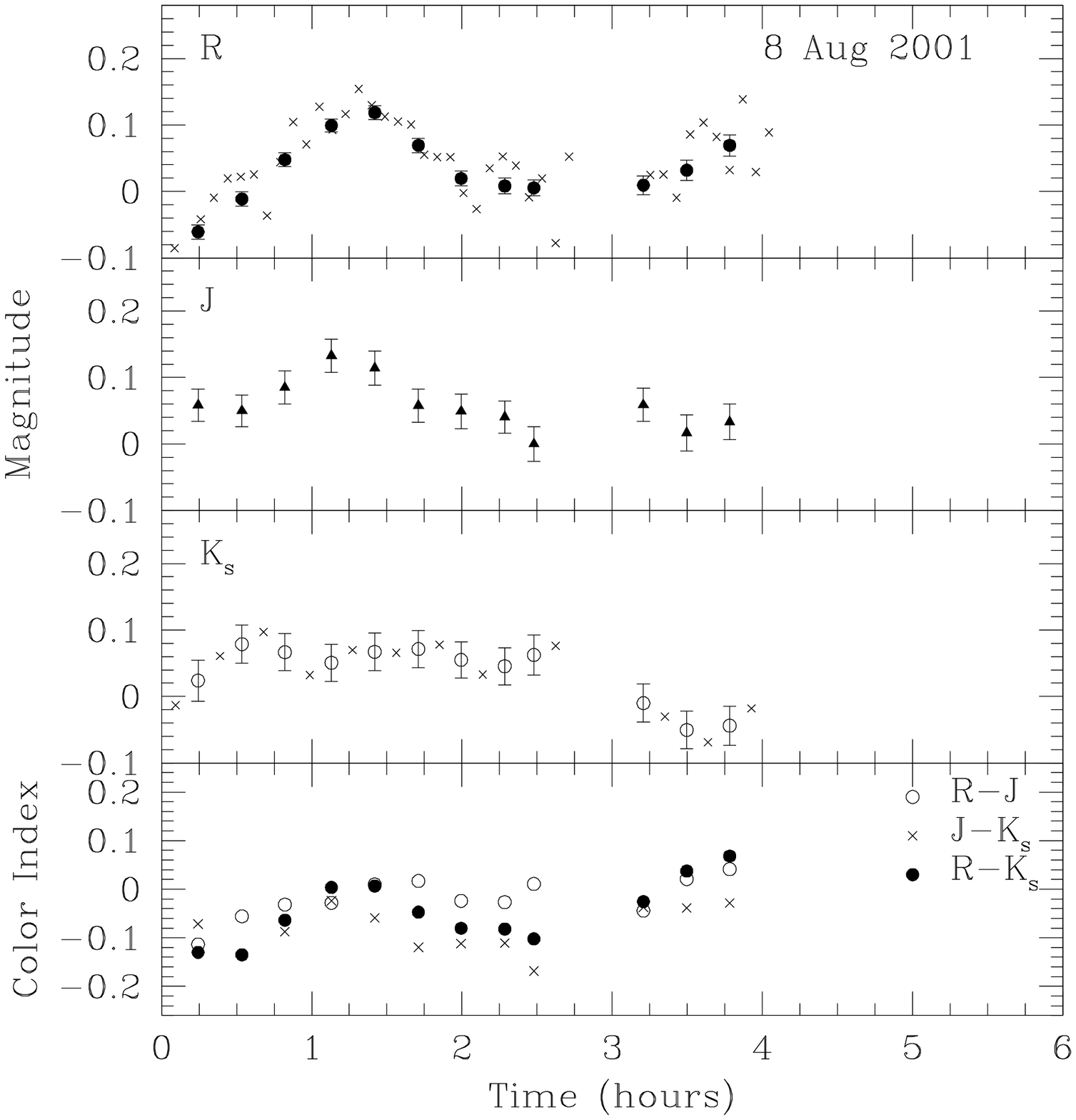}
\includegraphics[width=5cm]{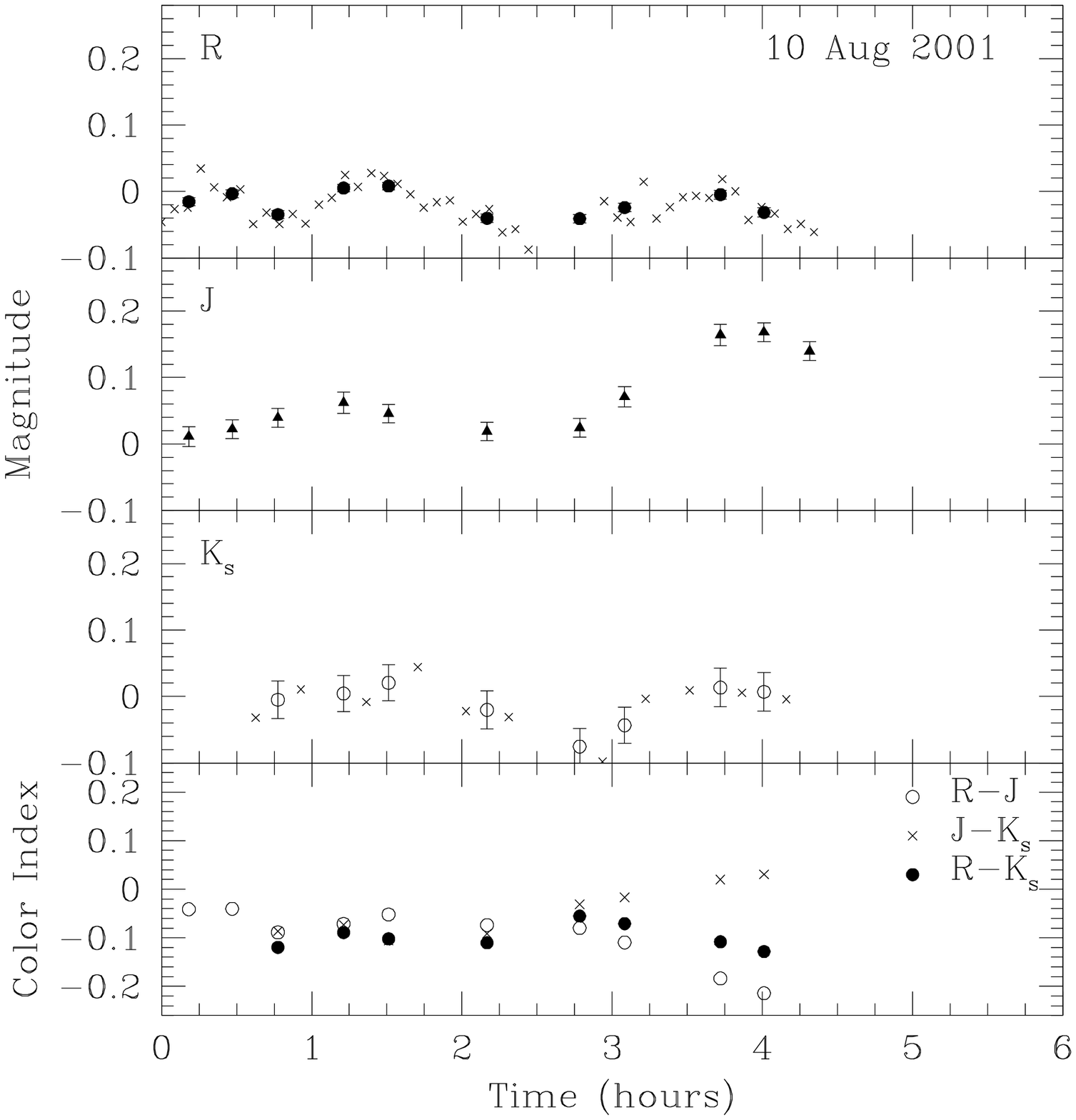}
\includegraphics[width=5cm]{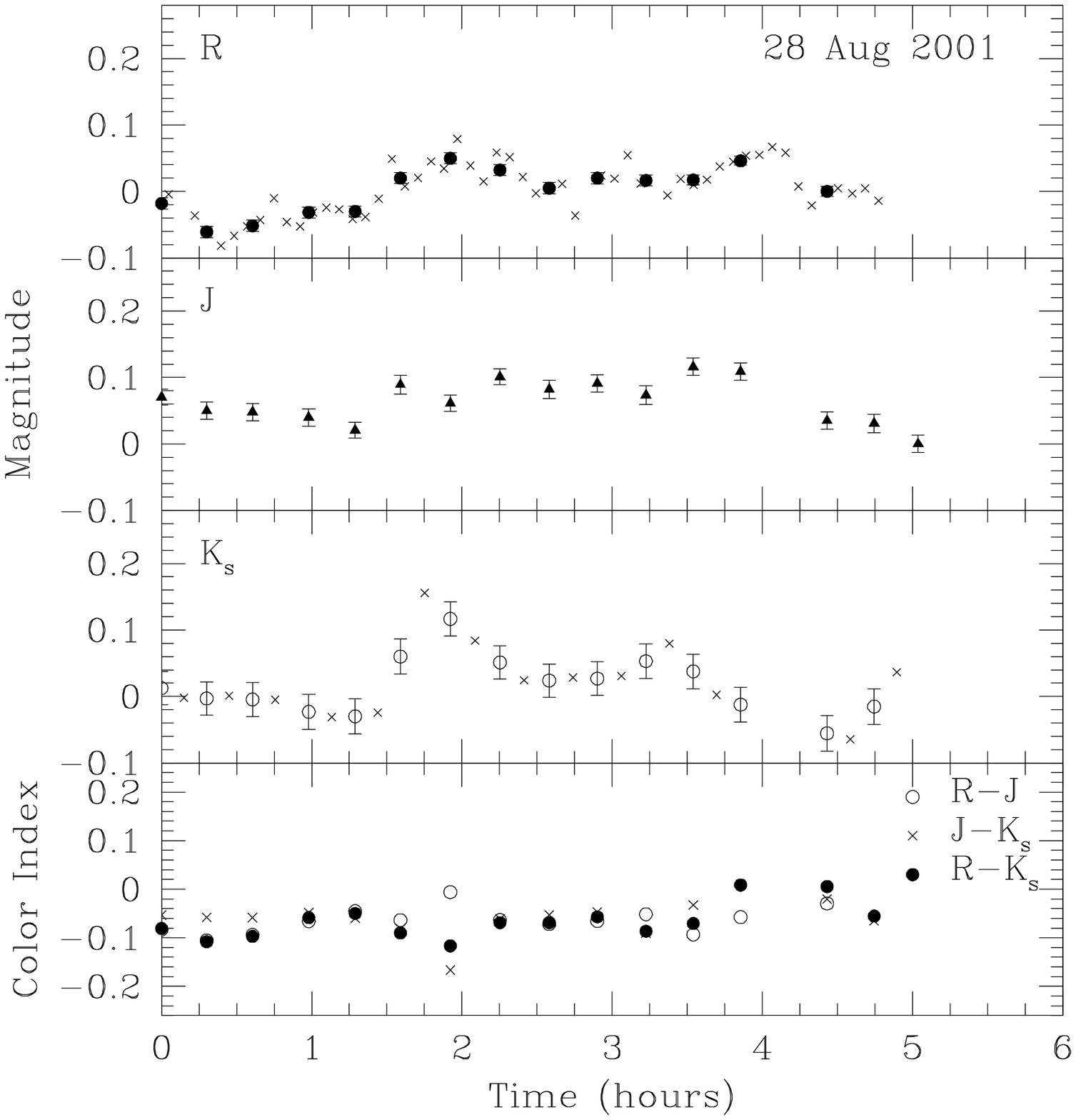}
\includegraphics[width=5cm]{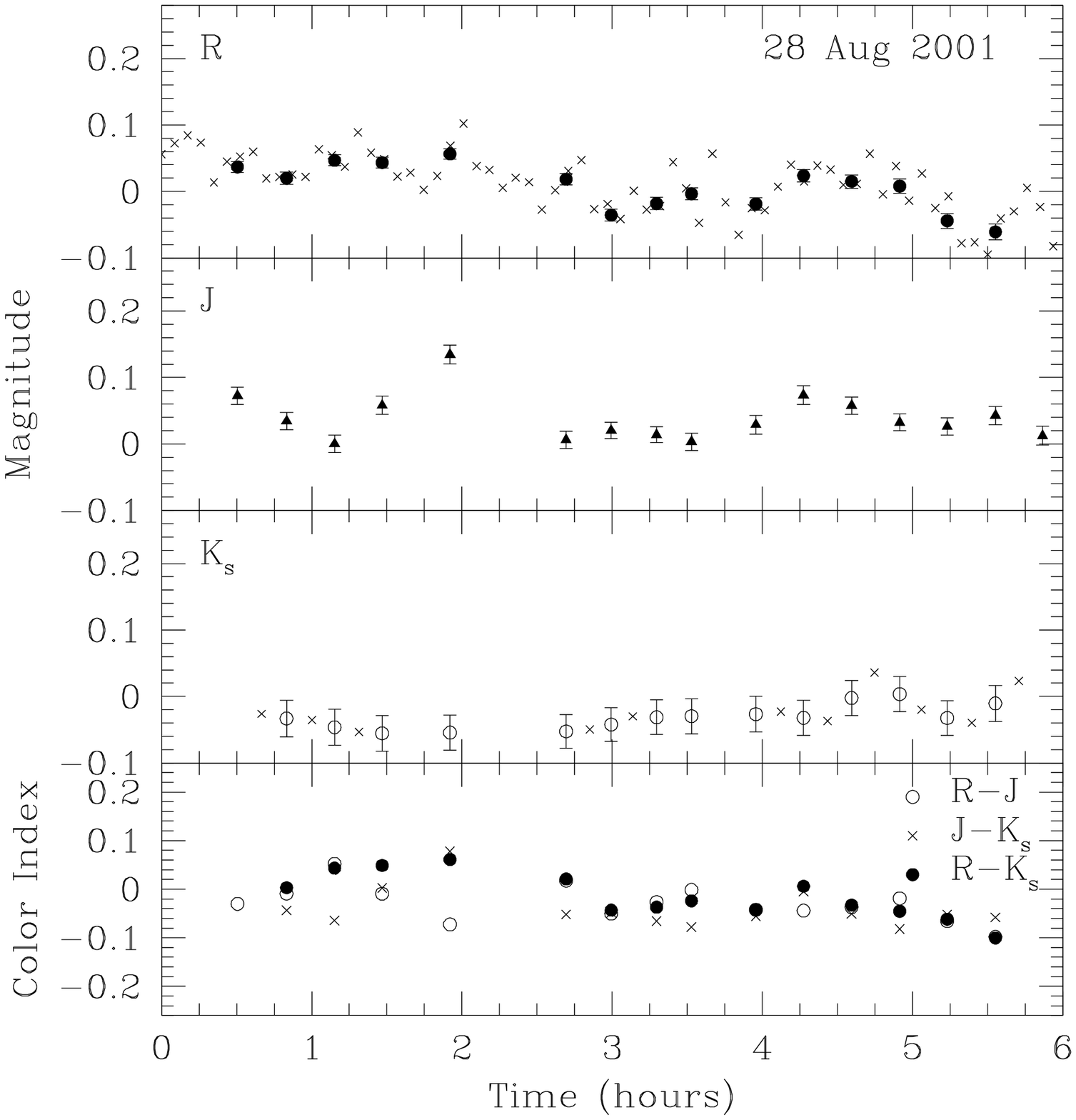}
\caption{From top  to bottom:  The $R$, $J$  and $K_s$  lightcurves of
V404  Cyg (after subtracting  the  ellipsoidal  modulation)  and the  color
indices $R-K_s$, $J-K_s$ and $R-J$. Crosses mark the original data and
circles the new resampled data.}
\label{curvaRIR}
\end{center}
\end{figure} 

\bsp


\end{document}